\documentclass[useAMS,usenatbib]{mn2e}

\usepackage{graphicx,amssymb,color}
\usepackage[normalem]{ulem}

\newcommand{\rev}{ }

\title[Lifetimes of WD discs]
{The lifetimes of planetary debris discs around white dwarfs}

\author[]{Dimitri Veras$^{1,2}$\thanks{E-mail: d.veras@warwick.ac.uk}\thanks{STFC Ernest Rutherford Fellow},
Kevin Heng$^{2,3}$
\\
$^{1}$Centre for Exoplanets and Habitability, University of Warwick, Coventry CV4 7AL, UK
\\
$^{2}$Department of Physics, University of Warwick, Coventry CV4 7AL, UK
\\
$^{3}$Center for Space and Habitability, University of Bern, Gesellschaftsstrasse 6, CH-3012 Bern, Switzerland
}

\pubyear{2020}

\begin{document}
\label{firstpage}
\pagerange{\pageref{firstpage}--\pageref{lastpage}}
\maketitle

\begin{abstract}
The lifetime of a planetary disc which orbits a white dwarf represents a crucial input parameter into evolutionary models of that system. Here we apply a purely analytical formalism to estimate lifetimes of the debris phase of these discs, before they are ground down into dust or are subject to sublimation from the white dwarf. We compute {\rev maximum} lifetimes for three different types of white dwarf discs, formed from (i) radiative YORP breakup of exo-asteroids along the giant branch phases at $2-100$~au, (ii) radiation-less {\rev spin-up} disruption of these minor planets at $\sim 1.5-4.5R_{\odot}$, and (iii) tidal disruption of minor or major planets within about $1.3R_{\odot}$. We display these {\rev maximum} lifetimes as a function of disc mass and extent, constituent planetesimal properties, and representative orbital excitations of eccentricity and inclination. We find that YORP discs with masses {\rev up to $10^{24}$~kg} live long enough to {\rev provide a reservoir of surviving cm-sized pebbles and} m- to km-sized boulders {\rev that can be perturbed intact} to white dwarfs {\rev with cooling ages of up to 10~Gyr}. Debris discs formed from {\rev the spin} or tidal disruption {\rev of these minor planets or major planets can} survive in a steady state for {\rev up to respectively 1~Myr or 0.01~Myr, although most tidal discs would leave a steady state within about 1~yr}. {\rev Our results illustrate that dust-less planetesimal transit detections are plausible}, and would provide particularly robust evolutionary constraints. Our formalism can easily be adapted to individual systems and future discoveries.
\end{abstract}

\begin{keywords}
Kuiper belt: general – minor planets, asteroids: general – planets and satellites: dynamical evolution and stability – stars: evolution – white dwarfs – protoplanetary discs.
\end{keywords}

\section{Introduction}

The death throes of planetary systems represent unique tracers of their evolutionary history. Giant branch host stars dynamically excite their planetary constituents, resulting in gravitational instabilities which eventually manifest themselves as detectable debris close to and within the photosphere of the eventual white dwarfs, where the chemical constituents of exo-planetesimals can be measured.

As a host star leaves the main sequence, it will expand its radius by a factor of hundreds, shed over half of its mass, and increase its luminosity by factor of thousands. All orbiting objects, including both major and minor planets, migrate outward due to the stellar mass loss \citep{omarov1962,hadjidemetriou1963,veretal2011}, but may fail to ``outrun'' the enlarging star and be engulfed \citep{kunetal2011,musvil2012,adablo2013,norspi2013,viletal2014,madetal2016,staetal2016,galetal2017,raoetal2018,sunetal2018,ronetal2020}. Exo-planetary analogues to Mars, Jupiter, Saturn, Uranus and Neptune are sufficiently separated from their parent stars to typically survive engulfment \citep{schcon2008,veras2016b}.

If the surviving major planets are sufficiently ``packed'' together, then the stellar mass loss could change their stability boundaries \citep{debsig2002,veretal2013,veretal2018,voyetal2013,musetal2014,vergan2015}. The stability boundaries between minor and major planets are also altered \citep{bonetal2011,debetal2012,frehan2014,petmun2017,musetal2018,smaetal2018}. The resulting gravitational instabilities are rarely triggered immediately, and usually only after the star has become a white dwarf. Although the minor planets themselves might survive engulfment, they struggle to survive intact because the highest luminosities achieved during the giant branch phases easily pushes around \citep{veretal2015a,veretal2019a} and spins up these bodies, commonly to breakup speed \citep{veretal2014b,versch2020}.

The subsequent significant reservoirs of pebbles, cobbles, boulders, minor planets and major planets which orbit a newly born white dwarf can reside at separations anywhere between a few au and on the order of $10^2$ au. This debris disc provides the primary source of mass that can be dynamically excited and perturbed towards the white dwarf\footnote{More distant sources provide minor contributions. Major planets are not expected to form beyond $10^2$ au. In the range $10^2 - 10^3$ au, large, Ceres-like minor planets largely survive and retain the same orbits that were set by the primordial stellar cluster evolution \citep{veretal2020b} but are not numerous. Boulders and pebbles could easily be radiatively thrust into this $10^2 - 10^3$ au region during the giant branch phases \citep{veretal2019a}, but represent a small total amount of mass. From $10^3 - 10^4$ au, inner exo-Oort clouds would survive \citep{stoetal2015,caihey2017}, but comets beyond $10^4$ au are thought to be severely depleted due to gravitational escape from giant branch mass loss \citep{verwya2012,veretal2014c,veretal2014d}.}.  If this disc survives in a largely collision-less state on Gyr timescales, then perturbed constituents would arrive at the white dwarf intact. Otherwise, the constituents might have already been pulverized into dust by the time they approach the white dwarf. One focus of our investigation here is to distinguish these possibilities by computing YORP disc lifetimes.

The exo-planetesimals which do reach the immediate region of the white dwarf, within several solar radii \footnote{Giant exoplanets themselves can also be perturbed towards the white dwarf, as evidenced by a reported ice giant planet orbiting at an approximate distance of just 0.07 au from WD J0914+1914 \citep{ganetal2019,verful2019,verful2020}.}, have now been observed in three successive stages of their evolution, as: 

\begin{enumerate}

\item Fully or partially intact, sometimes with accompanying fragments, orbiting the white dwarf \citep{vanetal2015,manetal2019,vanetal2019}, 

\item Broken up into an annulus of dust and sometimes gas around the white dwarf \citep{zucbec1987,farihi2016,manetal2020}, and

\item Chemically stratified at the atomic level in the photosphere of the white dwarf \citep{vanmaanen1917,vanmaanen1919,zucetal2007,dufetal2010,kleetal2010,ganetal2012,juryou2014}. 

\end{enumerate}

\noindent{}So far the first two stages have been observed concurrently, with dusty effluences and fragment distributions of at least one disintegrating exo-planetesimal that are detectable with photometric transit depths exceeding 50 per cent \citep{vanrap2018}. In the third stage, the white dwarf photosphere reveals exquisite chemical profiles of exo-planetary material at a level of detail exceeding that of any other exoplanetary detection technique \citep{haretal2018,holetal2018,doyetal2019,swaetal2019b,bonetal2020}.

Linking and understanding the timescales for all of these stages relies on first understanding the lifetime of the debris disc in the first stage. This evolution of this disc, which is the second focus of our study, sets up the subsequent evolution of the system, where dust and gas is generated, and eventually accreted onto the white dwarf.

The replenishment timescale -- how often exo-planetesimals approach, break-up around and deposit themselves into a white dwarf -- is one of the most important unknown parameters in post-main-sequence planetary science. Two reasons are

\begin{enumerate}

\item This replenishment timescale is linked to and perhaps constrained by the 25-50 per cent of the white dwarf population which is ``metal polluted'' (containing photospheric exo-planetary matter), the 1.5 per cent of white dwarfs which are both metal polluted and contain a dusty disc \citep{wiletal2019}, the 0.07 per cent of white dwarfs which are metal-polluted and contain both observable dust and gas in a surrounding disc \citep{manetal2020}, and the observational biases against their detection \citep[e.g.][]{bonetal2017}.

\item Helium-rich white dwarfs can retain accreted exo-planetary material in their convective zones for up to about one Myr, depending on the age of the star. Lower limits for this accumulated mass are detectable \citep{faretal2010,giretal2012,xujur2012}. Hence the replenishment timescale, coupled with measured limits on accumulated mass, can place constraints on the deposition or accretion rate onto the white dwarf.

\end{enumerate}

Previous estimates of the replenishment timescale \citep{giretal2012} mix inferred convection zone masses from helium-rich metal-polluted white dwarfs (DBZ spectral type) with accretion rates of hydrogen-rich metal-polluted white dwarfs (DAZ spectral type). This inconsistency motivates alternate approaches, despite having provided a useful, working estimate ($\sim 10^4-10^6$ yr) for nearly a decade.

From theoretical modelling perspectives, identifying the lifetime of different stages of a debris, dusty or gaseous disc has not been a particular focus. Instead, the much shorter formation process of a debris disc has received more attention. Further, for the au-scale debris discs which are produced from giant branch luminosity spin-up of exo-asteroids to breakup speed \citep{veretal2014b,versch2020} -- an effect known as the YORP effect \citep{rubincam2000} -- the authors did not consider evolution after formation. 

For the debris discs generated from exo-planetesimals approaching the white dwarf, there are (at least) two formation channels: tidal disruption within the Roche radius of the white dwarf \citep{graetal1990,jura2003,debetal2012,veretal2014a,malper2020a,malper2020b} and {\rev spin} disruption outside of the Roche radius \citep{makver2019,veretal2020a}. The {\rm spin} disruption mechanism {\rev occurs when the exchange between spin and angular momentum of exo-planetesimals on highly eccentric orbits becomes chaotic, and can spin these objects up to break-up speed. This mechanism} applies {\rev to} aspherical asteroids (most solar system asteroids are aspherical) whose spin rate is altered with each pericentre passage, and can eventually increase to break-up speed.

Most post-formation evolution models of the discs which are produced from tidal disruption have considered the combined effect of gas and dust, or gas only \citep{bocraf2011,rafikov2011a,rafikov2011b,metetal2012,rafgar2012,kenbro2017b,mirraf2018,ocolai2020}. But what of the evolution of the debris discs which are formed beyond the sublimation distance -- sometimes well beyond this distance -- and in particular for cold white dwarfs? For this gas-free case, \cite{kenbro2017a} investigated collisional cascades of the debris over $10^6$ yr using a sophisticated coagulation and fragmentation code, and incorporated a steady stream of external particles throughout the disc evolution.

\begin{figure*}
\includegraphics[width=16cm]{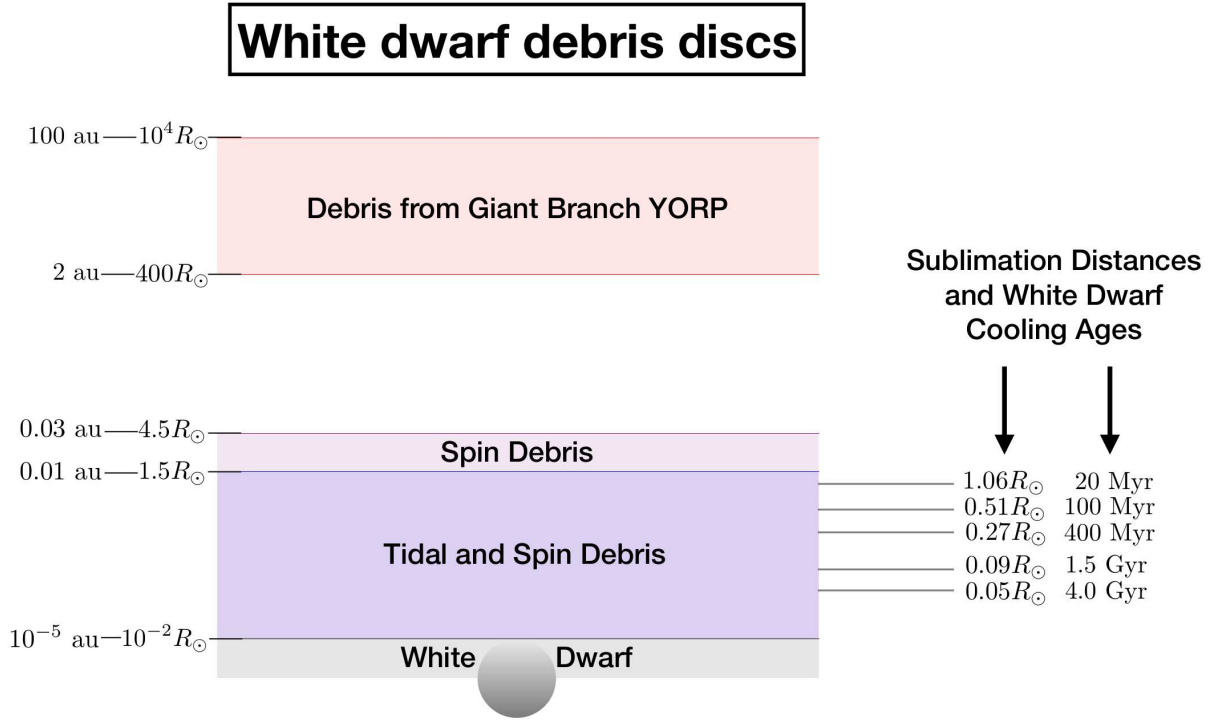}
\caption{
A cartoon with logarithmic vertical spacing which illustrates the predominant types and spatial extents of debris discs orbiting white dwarfs. Not included are exo-Oort clouds, which would be largely depleted, nor pockets of debris which could exist in stable resonant configurations with surviving major planets.  Exo-asteroids which avoid engulfment into the star during the giant branch phases would likely have arrived around the white dwarf at a separation of at least 2 au and in a fragmented state from the radiative YORP effect. Surviving fragments which are then perturbed towards the white dwarf (by surviving major planets) could be (further) broken up due to radiation-less {\rev spin} disruption or tidal disruption; $1.5R_{\odot}$ represents the approximate maximum tidal disruption separation for solid bodies. Given on the right are fiducial sublimation distances as a rough function of white dwarf cooling age; they indicate where a tidal debris disc can no longer be considered to be gas-free.
}
\label{cart}
\end{figure*}

Here, we instead apply a version of the entirely analytic formalism of \cite{hentre2010} to estimate the lifetime of three types of debris discs around white dwarfs: those produced by the YORP effect, those produced from radiation-less rotational disruption {\rev (henceforth labelled as ``spin'' discs)}, and those produced from tidal disruption; see Fig. \ref{cart} for a schematic overview. This formalism, while necessarily imposing some restrictive assumptions, allows us to explore the entire parameter space as a function of disc and planetesimal properties. {\rev The results can then be adapted to individual systems.}

In Section 2 we justify in detail our numerical choices for the input parameters to the model. Then in Section 3 we illustrate the output quantities from the model, and in some cases slightly generalize the formulae of \cite{hentre2010}. We perform the computations and present our results in Section 4 and then discuss these results in Section 5. We summarize our investigation in Section 6.

\section{Given variables} 

Required inputs into the model of \cite{hentre2010} include properties from the star, individual planetesimals, and the disc as a whole. We select parameter ranges for each of these variables based on a mixture of observational and theoretical constraints.

\subsection{The star: $M_{\star}$}

From the white dwarf itself, only the mass $M_{\star}$ is required to be input into the formalism of \cite{hentre2010} {\rev and in particular \cite{henmal2013}, who adopted an arbitrary stellar mass}. The vast majority of single white dwarfs harbour masses in the range $0.4-0.8 M_{\odot}$, with a peak around $0.60-0.65 M_{\odot}$; {\rev this peak is sharp} \citep{treetal2016,cumetal2018}.

From the mass alone, the radius of the white dwarf $R_{\star}$ can be analytically estimated. {\rev $R_{\star}$ can be expressed entirely in terms of $M_{\star}$ \citep{nauenberg1972,verrap1988}

\begin{equation}
\frac{R_{\star}}{R_{\odot}}
\approx
0.0127 
\left(  
\frac{M_{\star}}{M_{\odot}}
\right)^{-1/3}
\sqrt{
1 - 0.607
\left(  
\frac{M_{\star}}{M_{\odot}}
\right)^{4/3}
}
\end{equation}

\noindent{}}such that a range of $M_{\star} = 0.4-0.8 M_{\odot}$ corresponds to $R_{\star} = 0.0156 - 0.0125 R_{\odot}$, or $R_{\star} = 7.26 \times 10^{-5}$ - $5.83 \times 10^{-5}$ au.

Other stellar parameters which are important for context but are not necessarily input variables are the white dwarf's temperature $T_{\star}$, luminosity $L_{\star}$ and cooling age $t_{\rm cool}$. The cooling age is defined as the time since the white dwarf was born {(\rev not the time from the birth in the stellar cluster)}, and is relevant for this paper because disc lifetime might be a function of $t_{\rm cool}$ {\rev due to radiative effects}.  All three parameters are related non-trivially, and both $T_{\star} \approx 4\times 10^3 - 1\times 10^5$K and $L_{\star} \approx 10^{2}-10^{-5}L_{\odot}$ are functions of $t_{\rm cool}$ \citep{mestel1952,altetal2010,koester2013}. The function is steep: by the time the white dwarf has cooled to $t_{\rm cool} \approx 10$~Myr, $T_{\star}$ reduces from its peak value to a value of a few tens of thousands of K, when $L_{\star} \approx 0.1L_{\odot}$. The oldest known white dwarf planetary systems have $t_{\rm cool} \approx 8$~Gyr \citep{holetal2017}.

Overall, we adopt the range $M_{\star} = 0.4-0.8 M_{\odot}$.

\subsection{The planetesimals: $M_{\rm p}$, $R_{\rm p}$, $\rho_{\rm p}$}

Properties of the planetesimals are more uncertain. We assume that the planetesimals which comprise the disc are spheres and have masses, radii and densities which are denoted by $M_{\rm p}$, $R_{\rm p}$ and $\rho_{\rm p}$, only two of which {\rev need} to be specified. 

Observationally, constraints on these properties are minimal. {\rev Giant branch} YORP discs have not yet been observed, {\rev spin} discs might have been generated in just one known system \citep{vanetal2019}, and tidal discs just have an assumed lower radius grain limit of about 1.5 $\mu$m \citep{xuetal2018a}.  

In fact, the spherical assumption, {\rev which would have negated the YORP effect}, itself is probably inappropriate, given e.g. the complexity of the transiting signatures seen in the WD 1145+017 white dwarf planetary system \citep{vanetal2015} from, for example, \cite{ganetal2016}, \cite{garetal2017}, \cite{izqetal2018}, \cite{rapetal2018} and \cite{xuetal2019}. The size of the one large planetesimal orbiting within the disc around SDSS J1228+1040 \citep{manetal2019} is uncertain by a few orders of magnitude (with a radius range of about 4-600 km), and likely represents a core fragment of density $\rho_{\rm p} \gtrsim 7.7$ g/cm$^3$ with non-zero internal strength.

Theoretical models, however, have established other constraints. The highest possible mass of the progenitor of the planetesimals orbiting WD 1145+017 is thought to be about 10 per cent the mass of Ceres, such that $M_{\rm p} \lesssim 10^{20}$~kg \citep{rapetal2016,veretal2017a,guretal2017}. The density of this progenitor, however, is not necessarily uniform \citep{duvetal2020}, leaving $\rho_{\rm p}$ unconstrained. Indeed, a disc composed of solely iron-rich core fragments is not unreasonable \citep{manetal2019}, just as is a disc composed of only porous rubble piles. 

We can also look for theoretical constraints in studies of disc formation, whether it be from tidal disruption \citep{graetal1990,jura2003,debetal2012,veretal2014a,malper2020a,malper2020b} or rotational disruption \citep{makver2019,veretal2020a}. \cite{malper2020b} illustrated that tidal disruption can produce planetesimals ranging in size from their numerical resolution limit of a few km up to hundreds of km. The size distribution of fragments resulting from rotational disruption remains unclear, but would be a strong function of the number of fissions \citep{scheeres2018,versch2020}.

{\rev The size distribution in YORP discs depends on the number of YORP-induced fissions, as well as the internal strength, density and shapes of the planetesimals. \cite{versch2020} modelled planetesimals in the size range of 1 m to 10 km, but were non-committal about the size of the smallest monolithic constituents of progenitor asteroids. A wider range of planetesimals may comprise discs formed from the fragments of collisions amongst major planets. However, such collisions are expected to be rare (most gravitational instability events induce escape; see Appendix A of \citealt*{veretal2016a}) compared to YORP processes, which would be near-ubiquitous.}

Overall, we characterize the planetesimals by their radius $R_{\rm p}$ and density $\rho_{p}$, where $M_{\rm p}$ is then obtained trivially. We adopt the ranges $R_{\rm p} = 10^{-5} - 10^2$ km and $\rho_{\rm p} = 1-8$ g/cm$^3$.

\subsection{The disc: $r_{\rm i}$, $r_{\rm o}$, $M_{\rm disc}$, $\sigma_{e}$, $\sigma_{i}$}

As for the disc, we parametrize it with five variables: inner radius $r_{\rm i}$, outer radius $r_{\rm o}$, total mass $M_{\rm disc}$, and the root-mean-squared eccentricity $\sigma_e$ and inclination $\sigma_i$ of the ensemble of its constituents.

We assume that the disc is a gas-free circular annulus. For tidal and perhaps {\rev spin} discs, this structure may also be referred to as a ``ring'' because the spatial extent of the detected dust is usually comparable in range to that of some of Saturn's rings. However, the simulations of \cite{malper2020b} show that in fact tidal disruption can spread planetesimals over a much wider range, whereas the transiting signatures around ZTF J0139+5245 \citep{vanetal2019} suggest the presence of a much narrower ring of material. Although this particular ring is likely to be extremely eccentric (with $e \approx 0.97$), most dusty disc/ring structures are observed to be near circular \citep{rocetal2015,farihi2016}. In reality, the situation is more complicated, and the eccentricity may be a function of semimajor axis \citep{manetal2016a}, with intensity patterns in at least the gas surrounding SDSS J1228+1040 harbouring eccentricities of potentially several tenths \citep{manetal2019}.

We confine our annuli to the three classes of discs defined in Fig. \ref{cart}, which we henceforth refer to as ``YORP discs'', ``{\rev spin} discs'', and ``tidal discs''.  We now provide additional detail about their constraints.

\subsubsection{The inner radius $r_{\rm i}$}

For YORP discs, the minimum disc edge is set by the engulfment distance along the giant branch phases, which is a few au \citep{kunetal2011,musvil2012,adablo2013,norspi2013,viletal2014,madetal2016,staetal2016,galetal2017,raoetal2018,sunetal2018,ronetal2020}. Further, because YORP breakup may readily occur at several tens of au, we set the maximum value of $r_{\rm i}$ at 30 au. Hence for these discs we set $r_{\rm i} = 2 - 30$ au.

For {\rev spin} and tidal discs, given our gas-free assumption, the value of $r_{\rm i}$ is set by the sublimation distance from the white dwarf as a function of $t_{\rm cool}$, and hence $L_{\star}$ and $T_{\star}$; Fig. \ref{cart} provides some approximate values of the sublimation distance as a function of $t_{\rm cool}$. We denote $r_{\rm sub}$ as an idealized distance beyond which no material is sublimated; in fact, as indicated by \cite{jura2008} and Section 6.1.1 of \cite{veras2016a}, the sublimation rate of a planetesimal is better characterized as a continuous function of separation from the star.  Nevertheless, for our purposes, we adopt the useful basic prescription of \cite{rafikov2011b}

\begin{equation}
r_{\rm sub} = \frac{1}{2} R_{\star} \left( \frac{T_{\star}}{T_{\rm sub}} \right)^2
\end{equation}

\noindent{}where $T_{\rm sub}$ is the sublimation temperature of a particular substance. Table 1 of \cite{rafgar2012} show that $T_{\rm sub} = 1600, 2000, 2100, 2300, 2300, 2600$ K for, respectively, iron, CAIs, olivine, SiC, Al$_2$O$_3$ and graphite. 

The range of $r_{\rm sub}$ is extensive. The minimum possible value of $r_{\rm sub}$ is obtained by considering the minimum $T_{\star}$ of about 4000 K, the maximum $T_{\rm sub}$ corresponding to graphite, and a value of $R_{\star}$ corresponding to $M_{\star} = 0.8M_{\odot}$. Those yield min$(r_{\rm sub}) = 1.19 R_{\star}$. In contrast, for the youngest white dwarfs, max$(r_{\rm sub}) > 1.0R_{\odot}$. 

Overall, for tidal and {\rev spin} discs, we set $r_{\rm i} = 0.4R_{\star} - 1.0R_{\odot}$.

\subsubsection{The outer radius $r_{\rm o}$}

Observations are currently limited to tidal discs\footnote{However, \cite{suetal2007} provided tantalizing hints of a YORP disc extending many tens of au to over $10^2$ au.}, where $r_{\rm o} = 1.0-1.2 R_{\odot} \approx 4.6-5.5 \times 10^{-3}$ au \citep[e.g.][]{ganetal2006}. This distance is commonly thought to coincide with the Roche, or tidal disruption distance $r_{\rm Roche}$, although $r_{\rm Roche}$ can actually vary considerably depending on the material properties of the planetesimal.  For solid bodies, \cite{veretal2017a} illustrated that $r_{\rm Roche} \approx 0.5-1.5R_{\odot}$. Hence, for tidal discs, we adopt {\rev max$(r_{\rm o}) = 1.3 R_{\odot}$}.

The extent of {\rev spin} discs remains unclear. \cite{makver2019} demonstrated the concept of YORP-less spin-up to destruction due solely to close passages with a white dwarf, and \cite{veretal2020a} applied the idea to the ZTF J0139+5245 system, where the semimajor axis of the debris is just 0.42 au \citep{vanetal2019}. \cite{veretal2020a} found that rotational fission could occur at separations up to $3R_{\rm Roche}$ across the progenitor density range of $1-8$ g/cm$^3$. We adopt max$(r_{\rm o}) = 4.5 R_{\odot}$ here for these discs, but acknowledge that {\rev spin} discs may extend further pending future parameter space explorations.

Regarding YORP discs, observational searches for rocky debris beyond $\approx 1.0R_{\odot}$ have been carried out by \cite{xuetal2013} and \cite{faretal2014} for the white dwarf planetary systems GD 362 and G29-38, respectively. These searches extended to tens of au, with a correspondingly decreasing ability to constrain the mass. For example, around GD 362, at 5~au the dust mass was constrained to be no larger than about $10^{22}$ kg (more massive than Ceres). The constraints were better for G29-38; at 11 au, the dust mass was constrained to be no larger than about $3 \times 10^{21}$ kg. Hence, massive planetesimal discs out to tens of au may exist but remain undetected. For YORP discs, we hence adopt max$(r_{\rm o}) = 100$~au but acknowledge that they could extend further \citep{veretal2019a}.

All of the maximum values of $r_{\rm o}$ are subject to a condition on the width of the annulus; the analytical formalism of \cite{hentre2010} relies on a disc which is neither too extended nor narrow.  To quantify the restriction on the disc's extent, we use the $f_{m}$ {\rev parameter from \cite{hentre2010}}

\begin{equation}
f_m \equiv \frac{4\left(r_{\rm o} - r_{\rm i}\right)}{r_{\rm o} + r_{\rm i}}.
\end{equation}

\noindent{}In order for the formalism to hold, $f_m$ should be of order unity, meaning  $r_{\rm o}/r_{\rm i} \sim 5/3$. We will adopt the range $1/2 < f_m < 3$, which corresponds to $9/7 \le  r_{\rm o}/r_{\rm i} \le 7$.

\subsubsection{The disc mass $M_{\rm disc}$}

Planetary debris discs orbiting white dwarfs have masses which are poorly constrained observationally. YORP discs have not yet been observed. For tidal systems with transiting debris, the amount of dust may be estimated by assuming that a homogeneous rectangular or cylindrical dust cloud creates the transits \citep{ganetal2016,xuetal2018a,vanetal2019,veretal2020a}. For both WD 1145+017 and ZTF J0139+5245, this mass was estimated to be about $10^{14}$~kg. {\rev This value, however, is several orders of magnitude smaller than the disc mass that would be needed to circularize planetesimals in the WD 1145+017 disc \citep{ocolai2020}.}

Alternatively, if nearly every metal-polluted white dwarf harbours an orbiting debris disc which acts as a conduit to the accretion, then we can look to the accumulated mass over the last Myr or so in helium-rich (DBZ spectral type) white dwarf convection zones. This accumulated mass in these stars is $10^{16} - 10^{22}$ kg \citep{faretal2010,giretal2012,xujur2012,veras2016a}.

From a theoretical perspective, the lower limit on the disc's mass and surface density can be arbitrarily small. The upper limit is set by the extent of the surviving planetary material from the giant branch phases. One terrestrial planet {\rev (of mass $10^{24}$ kg)} could easily be kicked towards the white dwarf due to a larger planet \citep{veretal2016a} or a binary stellar companion \citep{bonver2015,hampor2016,petmun2017,steetal2017,steetal2018,veretal2017b}, disrupt and form a disc; such discs are even massive enough to trigger second-generation planetesimal formation \citep{schdre2014,voletal2014,vanetal2018}. However, major planets are not nearly as frequent as, say, $10^{22}$ kg asteroids within a belt \citep{bonetal2011,debetal2012,frehan2014,antver2016,antver2019,musetal2018,smaetal2018}, or as frequent as moons around one of the surviving planets \citep{payetal2016,payetal2017}.

The progenitor minor planets in YORP discs could result from a giant impact and subsequent breakup of a terrestrial {\rev planet ($\sim 10^{24}$ kg)}. Alternatively, the disc could be similar in mass to the solar system's Main Belt ($\sim 10^{21}$ kg), or a much less massive disc.

Overall then, we establish a disc mass range encompassing all of the above estimates: {\rev $M_{\rm disc} = 10^{12} - 10^{24}$ kg, or $1.7 \times 10^{-13} - 1.7 \times 10^{-1}M_{\oplus}$}.

\subsubsection{The dispersions $\sigma_{e}$ and $\sigma_{i}$}

Several white dwarf dusty and gaseous planetary discs showcase dynamical activity. However, a procedure for mapping this activity into a root mean square eccentricity ($\sigma_{e}$) and inclination ($\sigma_i$) of the {\rev planetesimals} (and then back to their progenitors) is not obvious. Nevertheless, we must set these variables as inputs, {\rev and have deliberately generalized the treatment of \cite{hentre2010} and \cite{henmal2013} so that $\sigma_{e}$ and $\sigma_{i}$ are not set at a fixed ratio.}

For systems with infrared excesses, the dynamical activity arises from changes in this flux over time \citep{xujur2014,faretal2018,xuetal2018b,swaetal2019a,wanetal2019,rogetal2020}. Although the more sudden changes could arise from a recent impact event (from perhaps an asteroid colliding with the extant disc; \citealt*{wanetal2019}), the flux changes more likely result from interactions amongst the dust itself and potentially with the gas (when it exists). Variability within the gaseous components \citep{wiletal2014,manetal2016b,redetal2017,cauetal2018,denetal2018} takes on a different form, but is not as directly related to our choices for $\sigma_e$ and $\sigma_i$ for solid debris.

All this activity arises from observed tidal discs (within $1.5 R_{\odot}$), where eccentricity and inclination changes are limited. {\rev As a dynamical comparison, these limitations do not exist in the aforementioned giant branch YORP discs. The planetesimals in those discs} can be travelling on orbits which encompass the entire range of eccentricity and inclination. The reason is because prior to break-up, the asteroid's orbit was likely altered through a ``supercharged'' radiative Yarkovsky effect, thousands of times more powerful than that in the solar system \citep{veretal2015a,veretal2019a}. 

Therefore, we do not uniformly adopt the assumption of \cite{hentre2010} that $\sigma_i = (1/2) \sigma_e$. This longstanding relation appeared to originate in the protoplanetary disc literature from both simulations \citep{cazetal1982} and analytical formulations \citep{horetal1985}. However, the formation channels of those discs are different than all of the ones considered in this paper, and those discs were often assumed near-coplanar and near-circular. 

Hence, in our computations, when $\sigma_e \ll 0.1$, we will often assume $\sigma_i = (1/2) \sigma_e$ as a default choice, but not always. In our analytical treatment, we leave $\sigma_e$ and $\sigma_i$ as free parameters which obey a Rayleigh distribution. $\sigma_e$ is naturally constrained by the boundaries of the disc, such that 

\begin{equation}
0 < \sigma_e \ll \frac{r_{\rm o} - r_{\rm i}}{r_{\rm o} + r_{\rm i}}.
\end{equation}

\noindent{}We assume that $\sigma_i$ is equivalently restricted {\rev such that}

\begin{equation}
0 < \sigma_i \ll \frac{r_{\rm o} - r_{\rm i}}{r_{\rm o} + r_{\rm i}}.
\end{equation}

\noindent{}{\rev This} restriction on the inclination is necessitated by the formalism but may not be reflective of the high and retrograde inclinations which are easily achieved in YORP discs \citep{veretal2015a,veretal2019a}.

\subsection{The end state of discs}

Eventually {\rev the spin and tidal debris are broken up into dust and gas and} accrete onto the white dwarf. So can the accretion process and rates help inform our initial conditions?

{\rev The interaction between this gas and} the dust is non-trivial and complicates the interpretation of the accretion rate \citep{metetal2012}. Further, if the {\rev dust is replenished or altered \citep{ocolai2020}} regularly from incoming debris, then the equilibrium condition of the {\rev dust} can alternate between high and low states, with or without the presence of gas \citep{kenbro2017a,kenbro2017b}. Further, incoming asteroids may bypass the {\rev spin or tidal discs} entirely and collide with the photosphere of the white dwarf \citep{broetal2017}. If the white dwarf is polluted simultaneously from the {\rev gas} and from ``direct hits'', then the accretion rate may contain both stochastic and steady elements \citep{wyaetal2014,turwya2020}. None of these complications provide sufficiently constrained predictions for us to alter our parameter choices.

\subsection{Parameter range summary}

We summarize the ranges for our initial parameters here, {\rev and note that we will perform a systematic march through the phase space. Unlike \cite{henmal2013}, we do not pursue a Monte Carlo approach.}

\begin{enumerate}

\item For all disc types, $M_{\star} = 0.4-0.8M_{\odot}$.

\item For all disc types, $R_{\rm p}  = 10^{-5}-10^{2}$ km. 

\item  For all disc types, $\rho_{\rm p} = 1-8$ g/cm$^3$.

\item For YORP discs, $r_{\rm i} = 2-30$ au. For tidal and {\rev spin} discs $r_{\rm i} = 0.4R_{\odot} - 1.5R_{\odot}$. 

\item Subject to the restriction $9/7 \le  r_{\rm o}/r_{\rm i} \le 7$: for YORP discs, max$(r_{\rm o}) = 100$ au; for {\rev spin} discs, max$(r_{\rm o}) = 4.5 R_{\odot}$; and for tidal discs, max$(r_{\rm o}) = 1.3 R_{\odot}$. 

\item For all disc types, $M_{\rm disc} = 10^{12} - 10^{24}$ kg. 

\item For all types of discs,

\begin{equation}
0 < \sigma_e \ll \frac{r_{\rm o} - r_{\rm i}}{r_{\rm o} + r_{\rm i}}.
\end{equation}

\item For all types of discs,

\begin{equation}
0 < \sigma_i \ll \frac{r_{\rm o} - r_{\rm i}}{r_{\rm o} + r_{\rm i}}.
\end{equation}



\end{enumerate}

\section{Disc lifetime}

The primary goal of this paper is to compute the {\rev maximum} lifetime, $t_{\rm disc}$, of three types of debris discs which orbit white dwarfs by using the analytical formalism of \cite{hentre2010} {\rev and \cite{henmal2013}}, thereby avoiding numerical integrations. {\rev As a first step, we define a disc as a structure which contains at least 10 planetesimals, such that the following condition must be satisfied}

\begin{equation}
R_{\rm p} \le \left( \frac{3 M_{\rm disc}}{20\pi\rho_{\rm p}} \right)^{1/3}
.
\label{disc10}
\end{equation}

\subsection{Initial auxiliary quantities}

In order to compute $t_{\rm disc}$, we first compute a series of relevant physical quantities. The first is the effective semimajor axis $a$ of the disc

\begin{equation}
a \equiv \frac{1}{2}\left(r_{\rm i} + r_{\rm o}\right).
\end{equation}

\noindent{}Then we can define the Hill radius, $R_{\rm Hill}$, of an individual planetesimal. The literature contains several expressions for $R_{\rm Hill}$, some of which include an eccentricity dependence \citep[e.g.][]{peawya2014} and others which refer to the mutual Hill radius between two objects \citep[e.g.][]{chaetal1996,steida2000}. For our purposes, we follow \cite{hentre2010} and define

\begin{equation}
R_{\rm Hill} \equiv a \left( \frac{M_{\rm p}}{3 M_{\star}} \right)^{\frac{1}{3}}.
\end{equation}

\noindent{}Because the same value of $a$ would be applied to each planetesimal rather than the instantaneous separation of that particular planetesimal, the formalism becomes less applicable as the disc's extent ($r_{\rm o} - r_{\rm i}$) is increased.

The height $h$ of the disc is given by

\begin{equation}
h = \frac{a\sigma_{i}}{\sqrt{2}}
\end{equation}

\noindent{}and the {\rev surface mass density $\Sigma$ is}

\begin{equation}
\Sigma = \frac{M_{\rm disc}}{\pi f_m a^2}
\end{equation}

\noindent{}from which follows the total mass density in the midplane 

\begin{equation}
\rho_0 = \frac{\Sigma}{\sqrt{\pi} a \sigma_i}
\end{equation}

\noindent{}as well as the number density in the midplane

\begin{equation}
n_0 = \frac{\Sigma}{\sqrt{2\pi} M_{\rm p} h}
.
\end{equation}



The mean motion of the disc, which is more commonly characterized as the orbital frequency $\Omega$, is

\begin{equation}
\Omega = \sqrt{\frac{GM_{\star}}{a^3}}
\end{equation}

\noindent{}and the radial velocity dispersion $\sigma_r$ is

\begin{equation}
\sigma_{r} = \frac{a \sigma_e \Omega}{\sqrt{2}}.
\end{equation}

With these variables, we can also define the Safronov number $\Theta$, which conveys the effect of gravitational focussing in scattering calculations

\begin{equation}
\Theta = \frac{G M_{\rm p}}{2 \sigma_{r}^2 R_{\rm p}}
,
\end{equation}



\noindent{}and a quantity $\Delta a$ which characterizes the typical radial separation between planetesimals, with

\begin{equation}
\Delta a = \frac{M_{\rm p}}{2 \pi \Sigma a}
.
\end{equation}

\subsection{Toomre stability}

{\rev With these quantities, we can determine if the disc is gravitationally stable to axisymmetric perturbations, a stability criterion known as Toomre stability \citep{toomre1964}. The disc is Toomre unstable} when either

\begin{equation}
M_{\rm p} < \frac{4\pi^3\Sigma^2a^4}{M_{\star}}
\ \ \ \ \ {\rm and} \ \ \ \ \
\sigma_{e} < \frac{\sqrt{2} \pi \Sigma a^2}{M_{\star}},
\end{equation}

\noindent{}or when

\[
M_{\rm p} \ge \frac{4\pi^3\Sigma^2a^4}{M_{\star}}
\ \ \ \ \  {\rm and} \ \ \ \ \
1 > \frac{M_{\rm p}  M_{\star}}
{\left(2 \pi \right)^3 \Sigma^2 a^4}
\]

\begin{equation}
\ \ \ \ {\rm and} \ \ \ \
\sigma_{e} < \sqrt{ \frac{M_{\rm p}}{\pi M_{\star}} 
\left[1 -  \frac{M_{\rm p}  M_{\star}}
{\left(2 \pi \right)^3 \Sigma^2 a^4} \right] }.
\end{equation}

{\rev We only compute maximum disc lifetimes for discs which are Toomre stable.}

\subsection{Disc classification}

\cite{hentre2010} made the distinction between ``cold'', ``warm'' and ``hot'' discs, depending on a number of factors. These distinctions crucially determine how $t_{\rm disc}$ is computed. Cold discs are discs where most of the planetesimal orbits do not cross, {\rev a condition which is equivalent to

\begin{equation}
\sigma_e < \frac{\left|\Delta a - 2R_{\rm p}\right|}{2a}
.
\end{equation}

\noindent{}In this case, the disc leaves the steady state only after the inherently chaotic nature of the multi-body problem generates a scattering event. We refer to this instability timescale as $t_{\rm chaos}$, which we estimate with the given parameters in the next subsection.}

{\rev In contrast,} both ``warm'' and ``hot'' discs feature collisions. If the collisions do not change the mass distribution, then the disc is ``warm''. Otherwise, the disc is ``hot''. {\rev Changing the mass distribution would entail any one of the following three events occurring:} (i) collisions generating gravitationally bound pairs, (ii) collisions {\rev excessively} eroding away the planetesimals, or (iii) the disc undergoing excessive viscous spreading. 

\begin{figure}
\includegraphics[width=9cm]{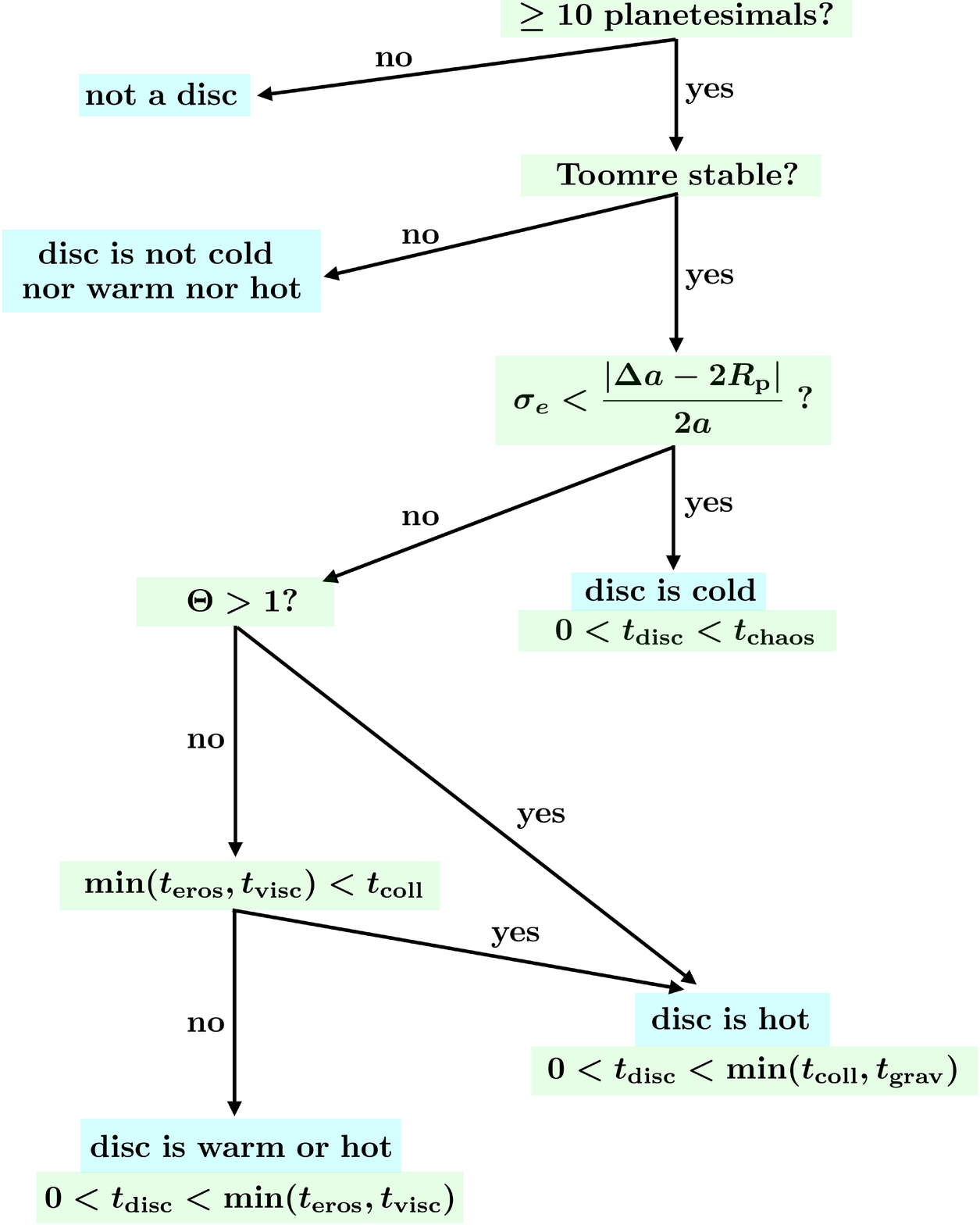}
\caption{
{\rev A flowchart describing how debris discs are classified and how disc lifetimes are bounded}.
}
\label{flow}
\end{figure}

{\rev These three events can all be parametrized. When $\Theta > 1$, then collisions can generate gravitationally bound pairs. Collisions can be thought of as gradually chipping away at the planetesimals on a timescale $t_{\rm eros}$, which we estimate with the given parameters in the next subsection. Viscous spreading can be assumed to occur on a timescale $t_{\rm visc}$, which we can again estimate. Hence, a warm disc requires, in addition to $\Theta \le 1$, that the collisional timescale $t_{\rm coll}$ must be shorter than $t_{\rm eros}$, $t_{\rm visc}$, and $t_{\rm disc}$. 

Because $t_{\rm disc}$ is the variable we seek, we cannot know a priori if $t_{\rm coll} < t_{\rm disc}$. Hence, with the given parameters, we can classify discs as ``cold'', ``hot'', or ``warm or hot''.  Further, in all three cases, we can place bounds on $t_{\rm disc}$. A final consideration is that gravitational scattering is not just limited to the ``cold'' disc case.  However, a collisional version of the gravitational scattering timescale is inherently different than $t_{\rm chaos}$, and is hence denoted as $t_{\rm grav}$. When $t_{\rm grav} < t_{\rm coll}$, then for hot discs $t_{\rm disc} < t_{\rm grav}$.  All of these conditions are summarized in Fig. \ref{flow}. }

\subsection{Timescale definitions}

{\rev We now explicitly estimate all of these timescales with the given variables.}

\subsubsection{Collisional timescale $t_{\rm coll}$}

\cite{hentre2010} {\rev developed an expression for the collision time $t_{\rm coll}$} by re-formulating the expressions from \cite{grelis1992} and \cite{dontre1993} in order to obtain a piecewise function depending on whether the system is in a dispersion-dominated or shear-dominated regime. {\rev Here, } we re-evaluate the expressions in Appendix A of \cite{hentre2010}, but now without assuming a fixed value of $\sigma_i/\sigma_e$. 

There are four cases, the first two being in the dispersion-dominated regime, and the final two in the shear-dominated regime.

\begin{enumerate}

\item When $\Theta < 1$ and $\sigma_r \ge \Omega \, {\rm max}\left(R_{\rm p}, \ 2^{\frac{1}{3}}R_{\rm Hill}\right)$, then

\begin{equation}
t_{\rm coll}^{-1} = 27.68 \frac{\sigma_r \sigma_i}{\sigma_e} n_0 R_{\rm p}^2
.
\label{dispreg}
\end{equation}

\item When $\Theta > 1$ and $\sigma_r \ge \Omega \, {\rm max}\left(R_{\rm p}, \ 2^{\frac{1}{3}}R_{\rm Hill}\right)$, then

\begin{equation}
t_{\rm coll}^{-1} = 30.49 \frac{\sigma_i}{\sigma_r \sigma_e} n_0 R_{\rm p} \Omega^2 R_{\rm Hill}^3
.
\end{equation}

\item When
 
$\Omega \sqrt{2^{\frac{4}{3}}R_{\rm p} R_{\rm Hill}} \le  \sigma_r \le \Omega \, {\rm max}\left(R_{\rm p}, \ 2^{\frac{1}{3}}R_{\rm Hill}\right)$, then

\begin{equation}
t_{\rm coll}^{-1} = 56.83 \frac{\sigma_i}{\sigma_e} n_0 R_{\rm p} \Omega R_{\rm Hill}^2
.
\end{equation}

\item When $\sigma_r \le \Omega\sqrt{2^{\frac{4}{3}}R_{\rm p} R_{\rm Hill}}$, then

\begin{equation}
t_{\rm coll}^{-1} = 32.44 \frac{\sigma_r \sigma_i}{\sigma_e} n_0 R_{\rm p}^{\frac{1}{2}}R_{\rm Hill}^{\frac{3}{2}}
.
\end{equation}

\end{enumerate}

\subsubsection{Viscous timescale $t_{\rm visc}$}

The viscous spread {\rev of the disc} through collisions is

\begin{equation}
\left(\frac{\delta a}{a}\right)_{\rm max} = 
\sigma_{r}
\sqrt{\frac
{32 a}
{G M_{\star}}
}
.
\end{equation}

{\rev \noindent{}Eq. (42) of \cite{hentre2010} then illustrated that the viscous timescale is

\begin{equation}
t_{\rm visc} = t_{\rm coll} \left(\frac{\delta a}{a}\right)_{\rm max}^{-2}
.
\end{equation}

}

\subsubsection{Erosion timescale $t_{\rm eros}$}

{\rev The erosion timescale $t_{\rm eros}$ is derived from Section 3.3 of \cite{hentre2010} as

\begin{equation}
t_{\rm eros} = t_{\rm coll} \left( \frac{Q_{\rm D}^{\star}}{\sigma_{r}^2} \right)
,
\end{equation}

\noindent{}where $Q_{D}^{\star}$ represents} the scaled disruption energy (or energy per unit mass required to disrupt a planetesimal into fragments). 

{\rev The value of} $Q_{D}^{\star}$ is composition- and size-dependent. Numerous empirical relationships have been developed between this energy and properties of colliding objects; for a recent summary, see \cite{gabetal2020}. Collisions of small bodies are said to be in the strength-dominated regime, whereas collisions of larger bodies are in the gravity-dominated regime. \cite{henmal2013} provided an expression which includes both regimes:

\begin{equation}
Q_{\rm D}^{\star} = 
Q_0 
\left[ \left(\frac{R_{\rm p}}{0.2 \ {\rm km}}\right)^{-0.4}  
+
\left(\frac{R_{\rm p}}{0.2 \ {\rm km}}\right)^{1.3}  
\right]
\label{Qeq}
\end{equation}

\noindent{}where the exponents were chosen to represent a mix of basalt and ice \citep{benasp1999}, which is a reasonable assumption for debris in white dwarf planetary systems. The constant $Q_0$ contains much of the variation that one might expect in $Q_{\rm D}^{\star}$ over a wide range of $R_{\rm p}$; we adopt a {\rev range of $Q_0 = 10^4 - 10^{7}$ erg/g}, just as in \cite{henmal2013}.

\subsubsection{Chaos timescale $t_{\rm chaos}$}

The mutual spacing within which a collection of {\rev collisionless} bodies would become unstable to gravitational perturbations is a longstanding problem in astrophysics, and one which is often not analytically tractable with more than three bodies.   

In planetary science, the discovery of extrasolar planets has inspired many stability applications which are focussed on major planets \citep{chaetal1996,chaetal2008,smilis2009,funetal2010,puwu2015}. For major planet masses, empirical numerical experiments have yielded functional formulae for $t_{\rm chaos}$ with constant coefficients that are actually a function of the number of the planets, the mass of the planets, and their orbital eccentricities and inclinations. This approach yields only approximate results, and with a lower mass limit $M_{\rm p}/M_{\star} \sim 10^{-10}$, which is unsuitable for the lowest planetesimal masses that we consider here (for 1 cm pebbles, $M_{\rm p}/M_{\star} \sim 10^{-33}$).

Therefore, we must find an expression for $t_{\rm chaos}$ as a function of $M_{\rm p}$. Equations 3-4 of \cite{zhoetal2007} provide such a relation. Written in our variables,

\begin{equation}
\log{\left( \frac{t_{\rm chaos}}{1 \ {\rm yr}} \right)} = A + U \log{\left( \frac{K_0}{2.3} \right)} 
\end{equation}

\noindent{}where

\begin{equation}
K_0 \approx 2^{2/3} \frac{a}{R_{\rm Hill}} \left( \frac{\Delta a}{2a + \Delta a} \right),
\end{equation}

\begin{equation}
A = -2 + \sigma_e \left(1 + \frac{2a}{\Delta a}  \right) - 0.27 \log{\left(\frac{M_{\rm p}}{M_{\star}}\right)},
\end{equation}

\noindent{}and

\[
U = 18.7 + 1.1 \log{\left(\frac{M_{\rm p}}{M_{\star}}\right)}
\]

\begin{equation}
\ \ \ \ \ - \left[16.8 + 1.2 \log{\left(\frac{M_{\rm p}}{M_{\star}}\right)} \right]   \sigma_e \left(1 + \frac{2a}{\Delta a} \right).
\end{equation}

\noindent{}Because these coefficients were derived for systems of 10 planets, one might question their applicability to the pebbles and boulders which we will sometimes consider. However, although these coefficients would change for fewer planets, \cite{funetal2010} indicated that for more than 10 planets, $t_{\rm chaos}$ becomes independent of the number of planets.

\subsubsection{Gravitational timescale $t_{\rm grav}$}

{\rev In order to compute the gravitational scattering timescale for discs with crossing planetesimal orbits $t_{\rm grav}$},
we use the formalism of \cite{steida2000}. First,

\begin{equation}
\left(t_{\rm grav}\right)^{-1} = \frac{d \ln{\left(\sigma_{e}^2\right)} }{dt}
                              = \frac{1}{\sigma_{e}^2} \frac{d\sigma_{e}^2}{dt}.
\end{equation}

\noindent{}Then, with Eq. (3.29) of \cite{steida2000} and with our assumption of equal-mass planetesimals,

\begin{equation}
\frac{d\sigma_{e}^2}{dt} = \frac{\Omega \Sigma a^2}{4 M_{\rm p}}\left\langle P_{\rm vs} \right\rangle
\end{equation}

\noindent{}where (their Appendix B)

\begin{equation}
\left\langle P_{\rm vs} \right\rangle = 
      \frac{16 M_{\rm p}^2\beta B}{\pi M_{\star}^2 \sigma_e \sigma_i}
                                       \int_{0}^{1}
\left[
\frac{
5 \mathcal{K}(\kappa)
-
\frac{12 \left(1 - \lambda^2 \right) \mathcal{E}(\kappa)}
        {1 + 3 \lambda^2}
}
{
\beta^2 + \left(1 - \beta^2\right) \lambda^2
}
\right]
d\lambda.
\label{PVS}
\end{equation}

In Eq. (\ref{PVS}),  $\beta = \sigma_{i}/\sigma_{e}$, and $\mathcal{K}(\kappa)$ and $\mathcal{E}(\kappa)$ are, respectively, complete elliptic integrals of the first and second kind, with argument 

\begin{equation}
\kappa = \sqrt{\frac{3}{4} \left(1 - \lambda^2\right) }
.
\end{equation}



\noindent{}The variable $B$ an be expressed through Eqs. (2.17) and (6.6-6.8) of \cite{steida2000} as

\begin{equation}
B = \log{\left[ \frac{\Lambda^2 + 1}{\Lambda_{\rm c}^2 + 1} \right]}
 - \frac{1}{\Lambda_{\rm c}^2 + 1} + \frac{1}{\Lambda^2 + 1}
\end{equation}

\noindent{}with

\begin{equation}
\Lambda = \frac{M_{\star}}{M_{\rm p}}\left(\sigma_{e}^2 + \sigma_{i}^2 \right)
\left(\sqrt{2} \sigma_{i} + \frac{2^{1/3}R_{\rm Hill}}{a}  \right)
\end{equation}

\noindent{}and

\[
\Lambda_{\rm c} = \left(\frac{M_{\star}}{M_{\rm p}}\right)
\left(\frac{2R_{\rm p}}{a} \right)
\left(\sigma_{e}^2 + \sigma_{i}^2 \right)
\]

\begin{equation}
\ \ \ \ \ \ \times
\sqrt{1+
\frac
{M_{\rm p} a}
{R_{\rm p} M_{\star} \left[\sigma_{e}^2 + \sigma_{i}^2 + 2^{-\frac{1}{3}}\left(\frac{R_{\rm Hill}}{a}\right)^2  \right]}
}.
\end{equation}

\subsection{Radiation drag}

{\rev One physical process that may act on shorter timescales than collisions, erosion, scattering or viscous spreading is radiation drag from the white dwarf. For most white dwarfs, such drag is negligible compared to the drag generated during the highly luminous giant branch phases} \citep{bonwya2010,donetal2010,veretal2015a,veretal2019a,zotver2020}. {\rev However, young white dwarfs are particularly luminous ($>10^{-1} L_{\odot}$), albeit briefly (for under 10 Myr). This high luminosity} may be important to consider \citep{veretal2015b}, particularly when the Yarkovsky effect is active for young, hot white dwarfs like WD J0914+1914 \citep{veras2020}.

We can roughly estimate the timescale for a planetesimal to be dragged through about half of the disc due to radiation. Whether or not the Yarkovsky effect is active can make a significant difference to the {\rev radiation drag} timescale (of three to four orders-of-magnitude), and depends on the shape, spin and internal density distribution of the planetesimals. Hence, here we provide the limiting maximum drag due to the Yarkovsky effect (Eq. 103 from \citealt*{veretal2015a}), as well as the residual drag -- which is the Poynting-Robertson drag (Eq. 111 from \citealt*{veretal2015a})  -- when the Yarkovsky effect is turned off.

\begin{equation}
t_{\rm PRdrag} = \left( \frac{r_{\rm o} - r_{\rm i}}{2} \right)
                     \left( \frac{da}{dt} \right)_{\rm PR}^{-1}
                  \approx \frac{\left(r_{\rm o} - r_{\rm i} \right) c^2 M_{\rm p} a}{R_{\rm p}^2 L_{\star}},
\label{PReq}
\end{equation}

\[
t_{\rm maxYark} = \left( \frac{r_{\rm o} - r_{\rm i}}{2} \right)
                     \left( \frac{da}{dt} \right)_{\rm maxYark}^{-1}
\]

\begin{equation}
\ \ \ \ \ \ \ \ \ \ \ \
   \approx \frac{2\left(r_{\rm o} - r_{\rm i} \right) c M_{\rm p} \sqrt{G M_{\star} a}}{R_{\rm p}^2 L_{\star}},
\label{Yarkeq}
\end{equation}

In Eqs. (\ref{PReq}) and (\ref{Yarkeq}), $c$ is the speed of light. {\rev For every case studied in this paper,} we {\rev conservatively} compute both $t_{\rm PRdrag}$ and $t_{\rm maxYark}$ by assuming a relatively high white dwarf luminosity of $0.1L_{\odot}$ (corresponding roughly to a cooling age of 10 Myr). {\rev Hence, if both $t_{\rm maxYark} < t_{\rm disc}$ and $t_{\rm maxYark} < 10$ Myr, then we flag the disc lifetime as {\it possibly} being dominated by radiation. Depending on the shape and physical properties of the planetesimal, the actual Yarkovsky timescale may be much longer.}

{\rev The much weaker Poynting-Robertson drag ensures that $t_{\rm PRdrag} > t_{\rm maxYark}$, but is not subject to the same level of uncertainty as Yarkovsky drag (which might never activate, or cancel itself through self-regulation). Hence, if both $t_{\rm maxYark} < t_{\rm disc}$ and $t_{\rm PRdrag} < 10$ Myr, then the disc's evolution {\it is definitely} dominated by radiation drag (but only for the high luminosity we sample of $10^{-1}L_{\odot}$ and assuming $t_{\rm disc}=$ max($t_{\rm disc}$)). As the results in the next section will show, only a small fraction of the discs that we investigate might be dominated by radiation drag, and only a minority of those are definitely dominated by radiation drag.}

\section{Results}

\begin{figure*}
\centerline{ {\bf \LARGE \underline{YORP debris discs from 2-3 au}} }
\centerline{}
\centerline{
\includegraphics[width=9.5cm]{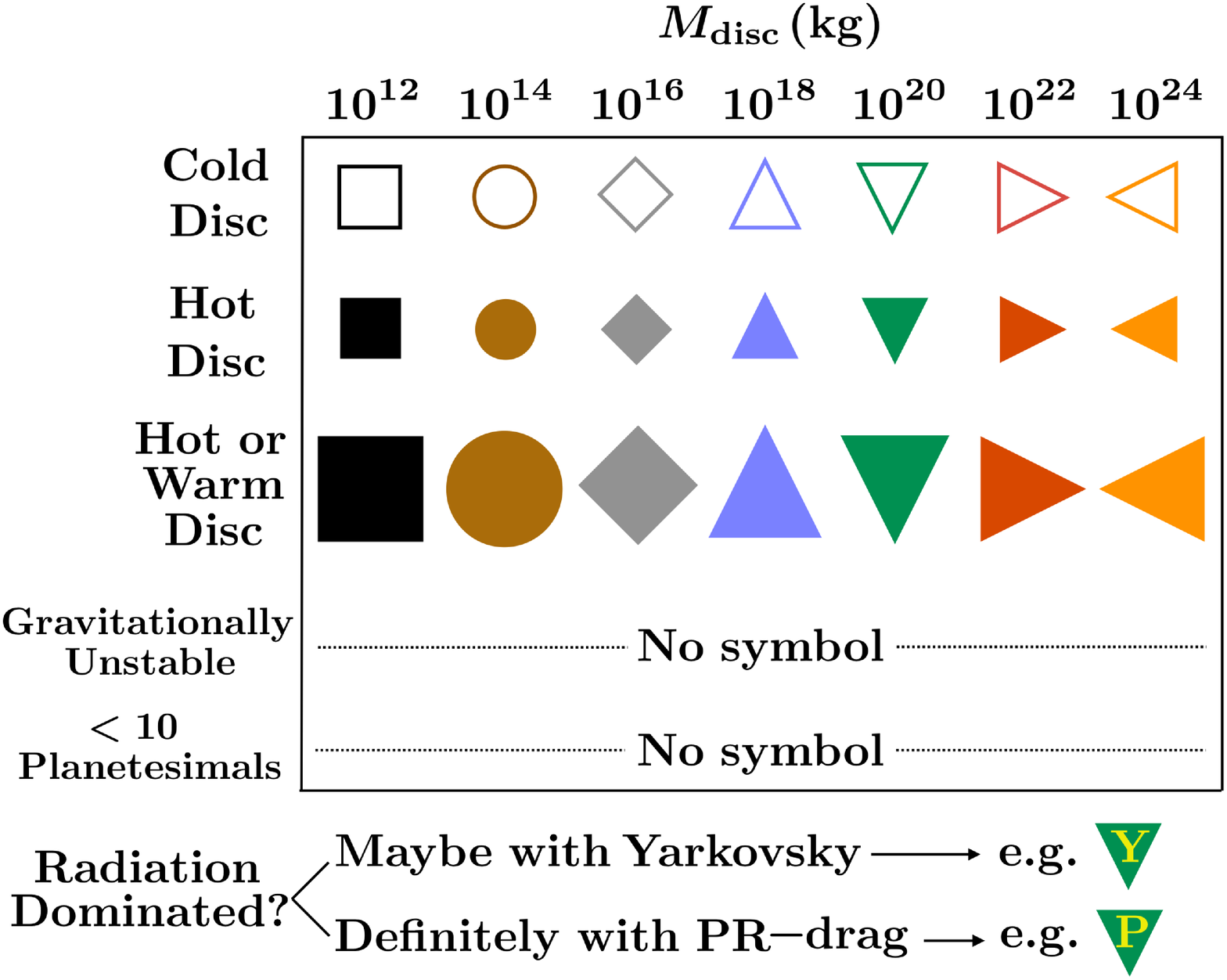}
}
\centerline{}
\centerline{
\includegraphics[width=9.5cm]{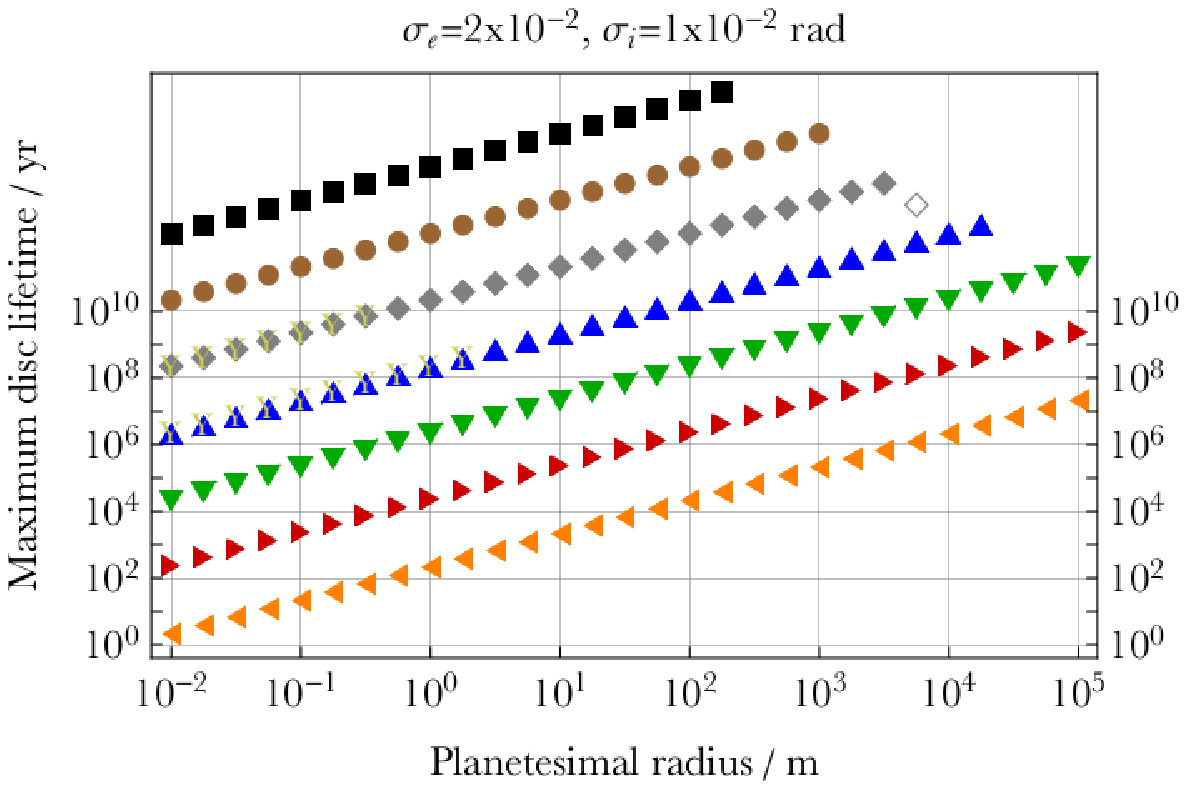}
\ \ \ \ \ \
\includegraphics[width=9.5cm]{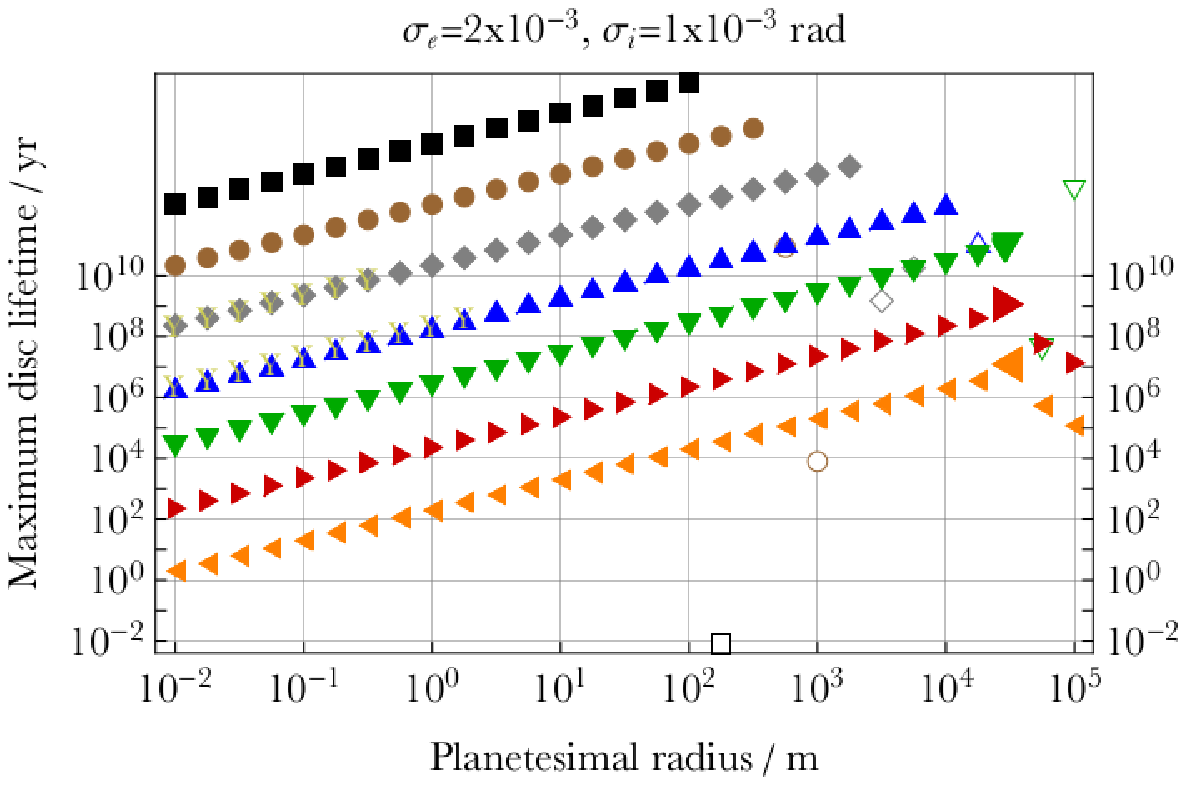}
}
\centerline{}
\centerline{
\includegraphics[width=9.5cm]{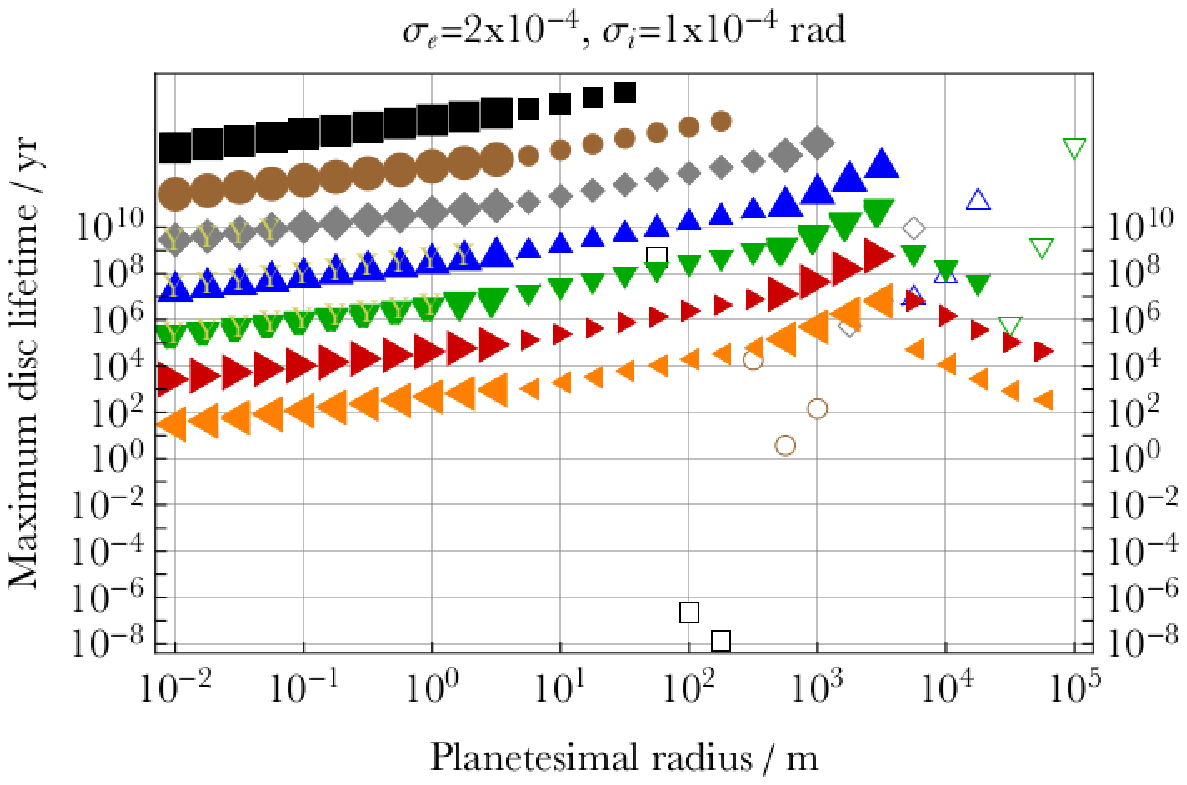}
\ \ \ \ \ \
\includegraphics[width=9.5cm]{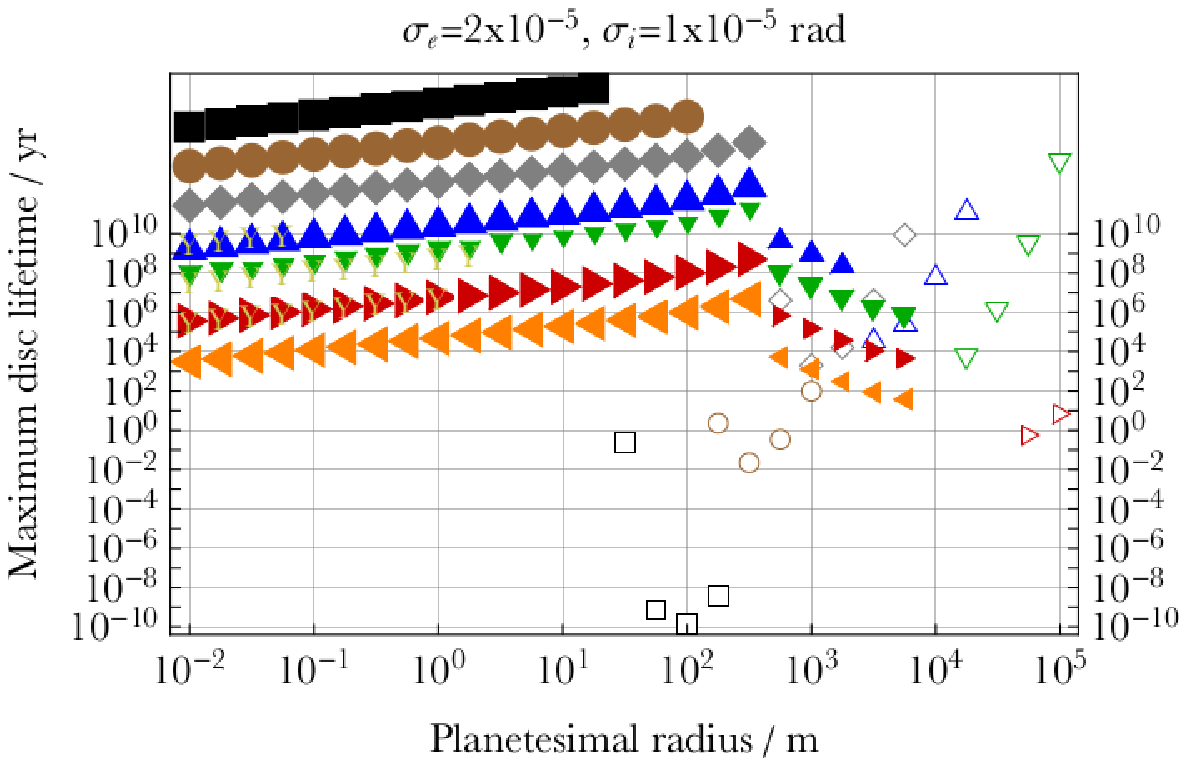}
}
\centerline{}
\caption{
{\rev Maximum debris disc lifetimes of narrow and close ``Main Belt''-like YORP discs (formed from YORP break up during the giant branch phases). For all plots, $\rho_{\rm p}=2$~g/cm$^3$, $M_{\star} = 0.6M_{\odot}$, $Q_{0} = 10^4$ erg/g, $r_{\rm i} = 2$ au and $r_{\rm o} = 3$ au. The $y$-axis labels do not exceed $10^{10}$ yr because any symbol above that value indicates a lifetime which exceeds the (approximate) age of the universe. For all plots, discs for which radiation is important (Y and P symbols) always assume a young luminous white dwarf with a constant luminosity of $0.1L_{\odot}$. The plots illustrate how as the disc becomes flatter and circular, the maximum lifetimes increasingly deviate from clear patterns. Overall, close and narrow YORP discs as massive as about $10^{20}$~kg can survive long enough in a steady state to regularly provide a source of intact planetesimals to potentially be perturbed towards the white dwarf.}
}
\label{YORP1}
\end{figure*}

\begin{figure*}
\centerline{ {\bf \LARGE \underline{YORP debris discs from 30-100 au}} }
\centerline{}
\centerline{}
\centerline{}
\centerline{
\includegraphics[width=9.5cm]{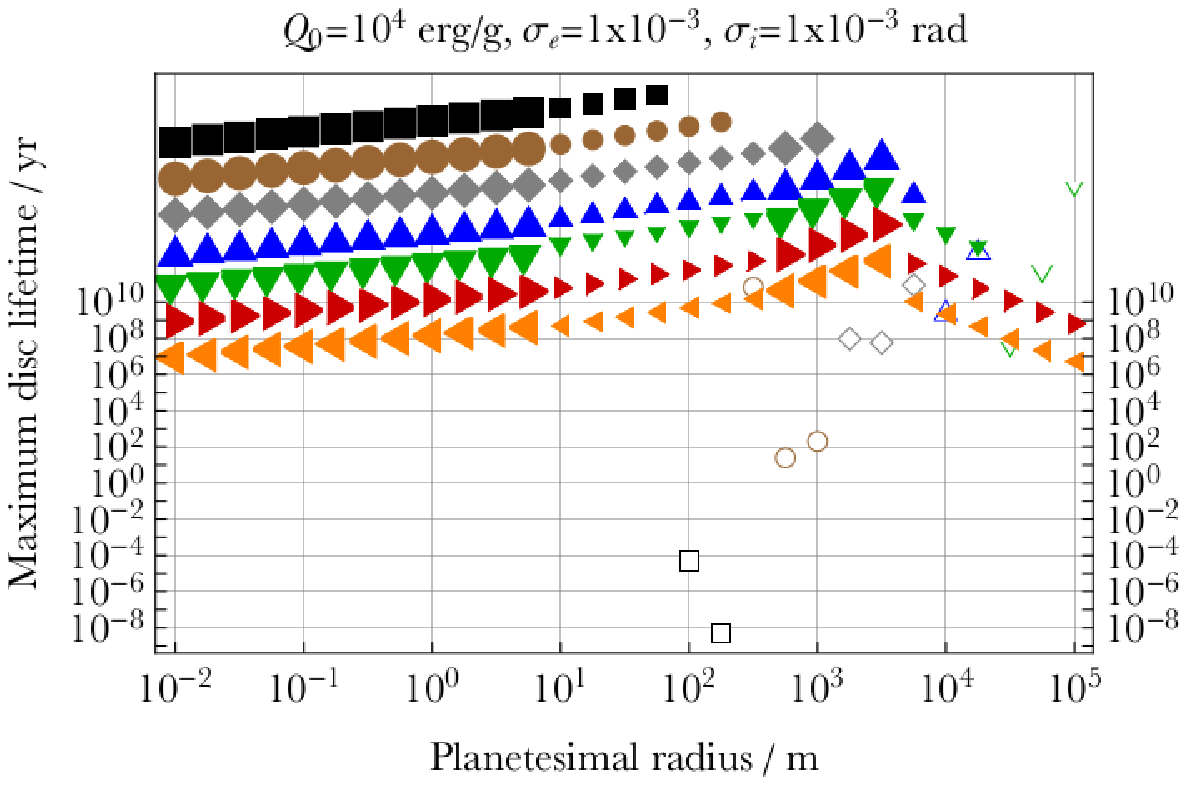}
\ \ \ \ \ \
\includegraphics[width=9.5cm]{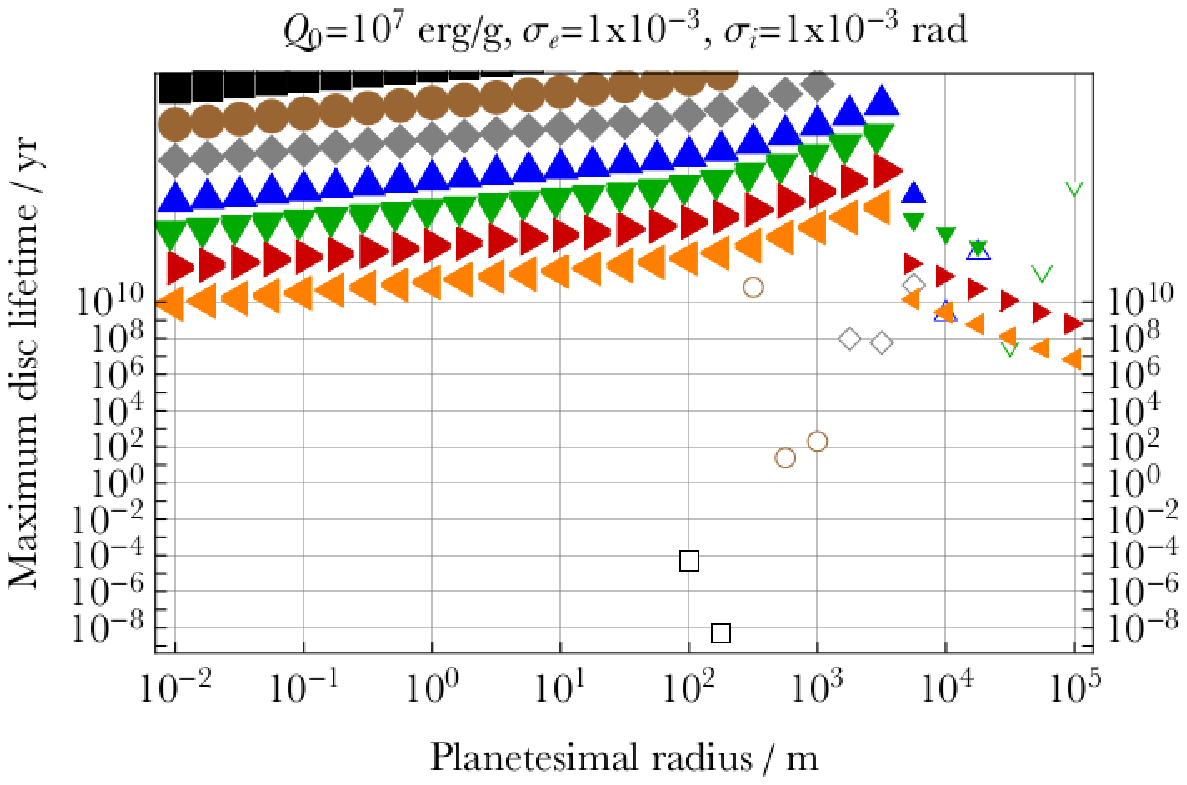}
}
\centerline{}
\centerline{}
\centerline{
\includegraphics[width=9.5cm]{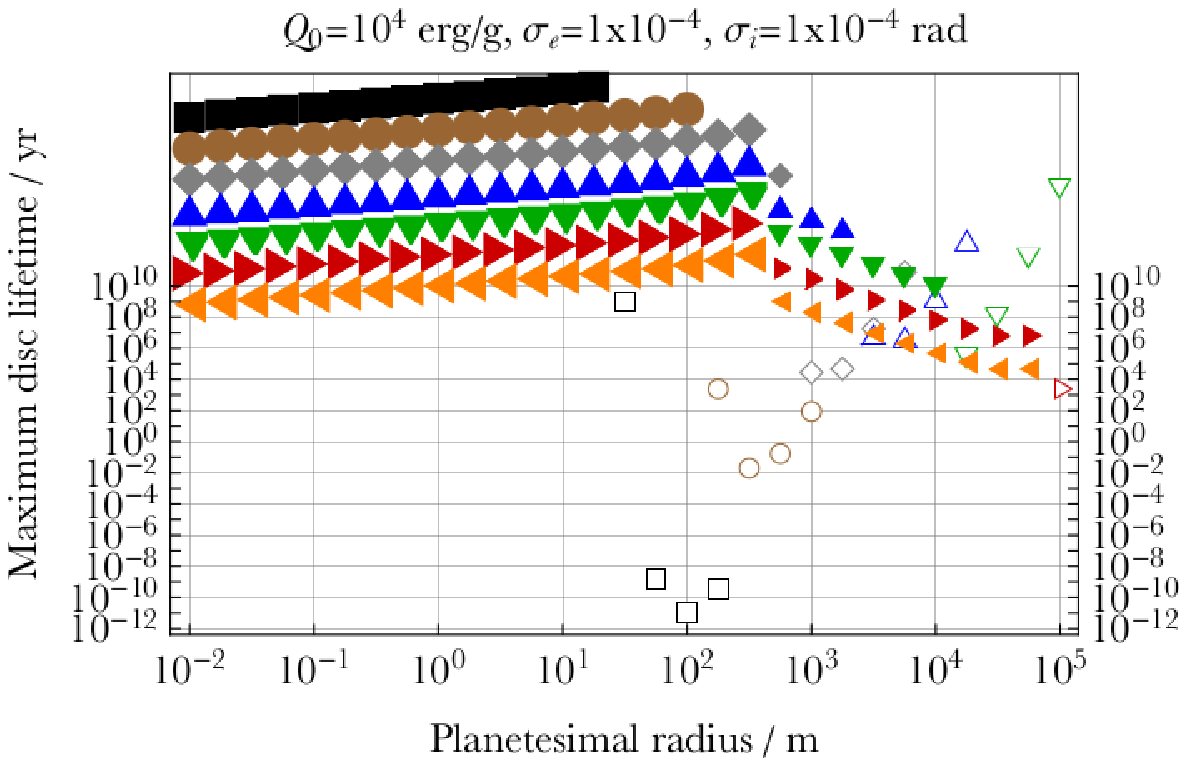}
\ \ \ \ \ \
\includegraphics[width=9.5cm]{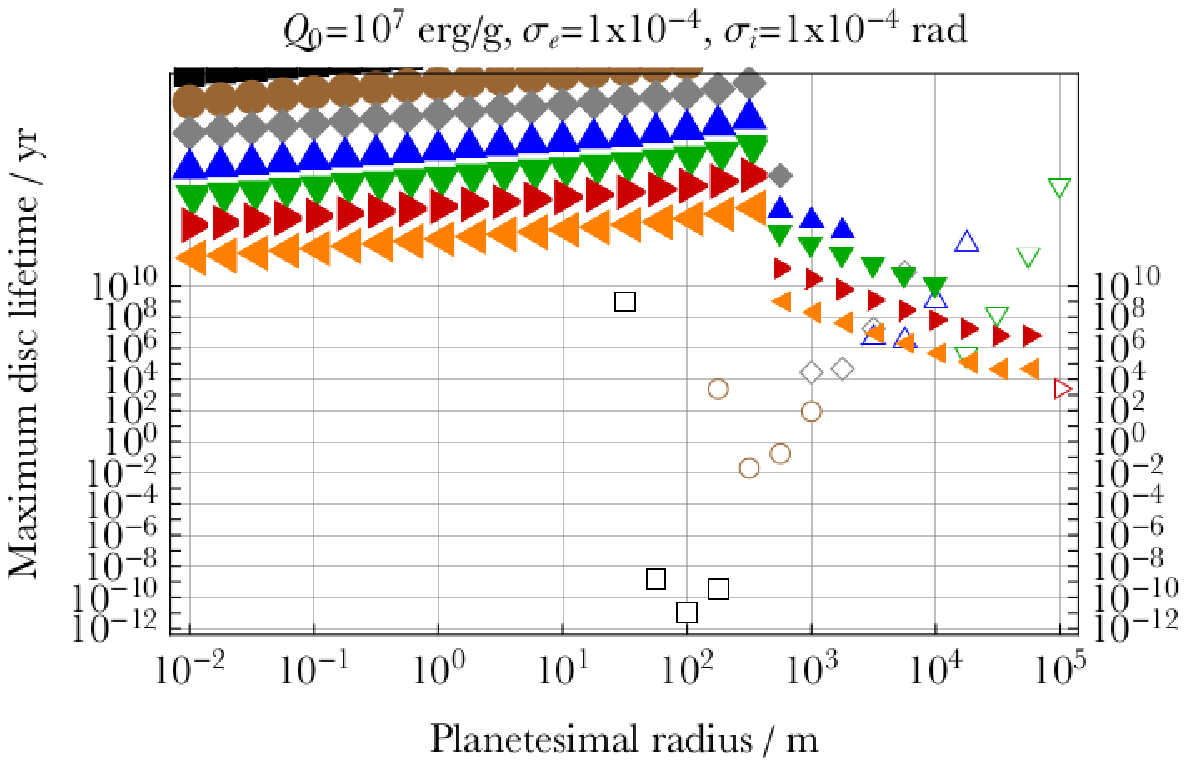}
}
\centerline{}
\centerline{}
\centerline{
\includegraphics[width=9.5cm]{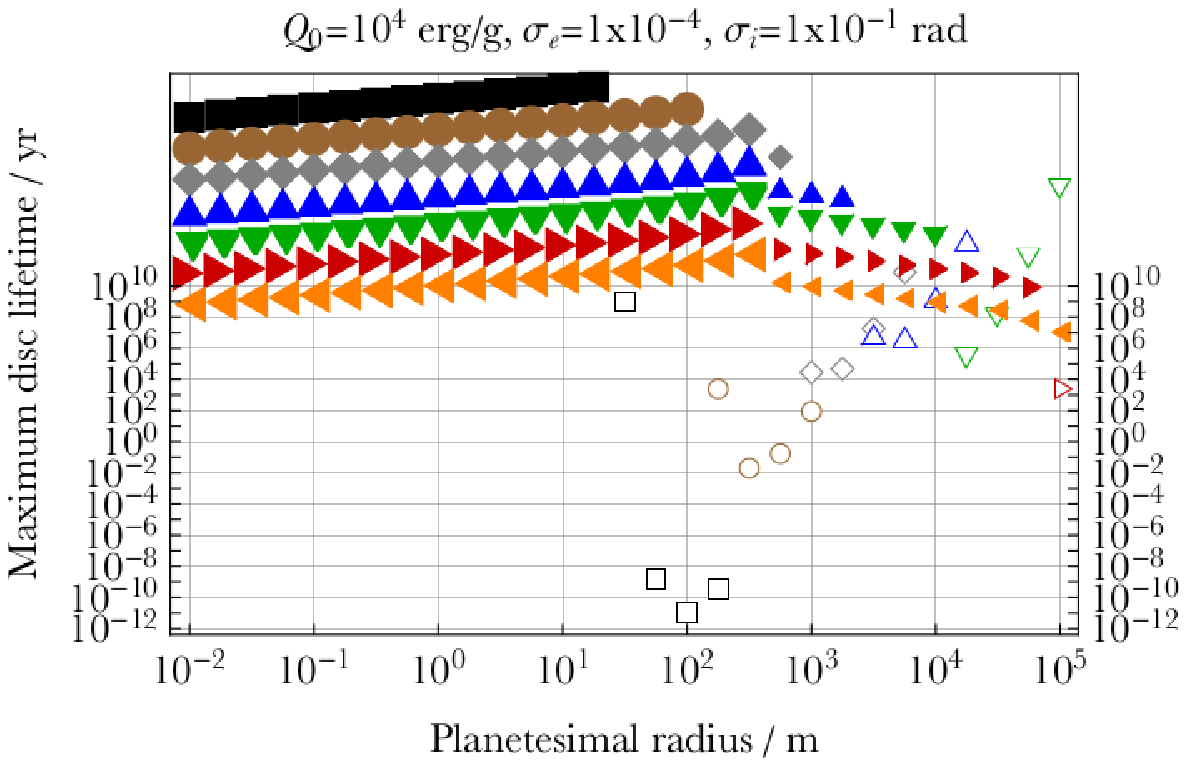}
\ \ \ \ \ \
\includegraphics[width=9.5cm]{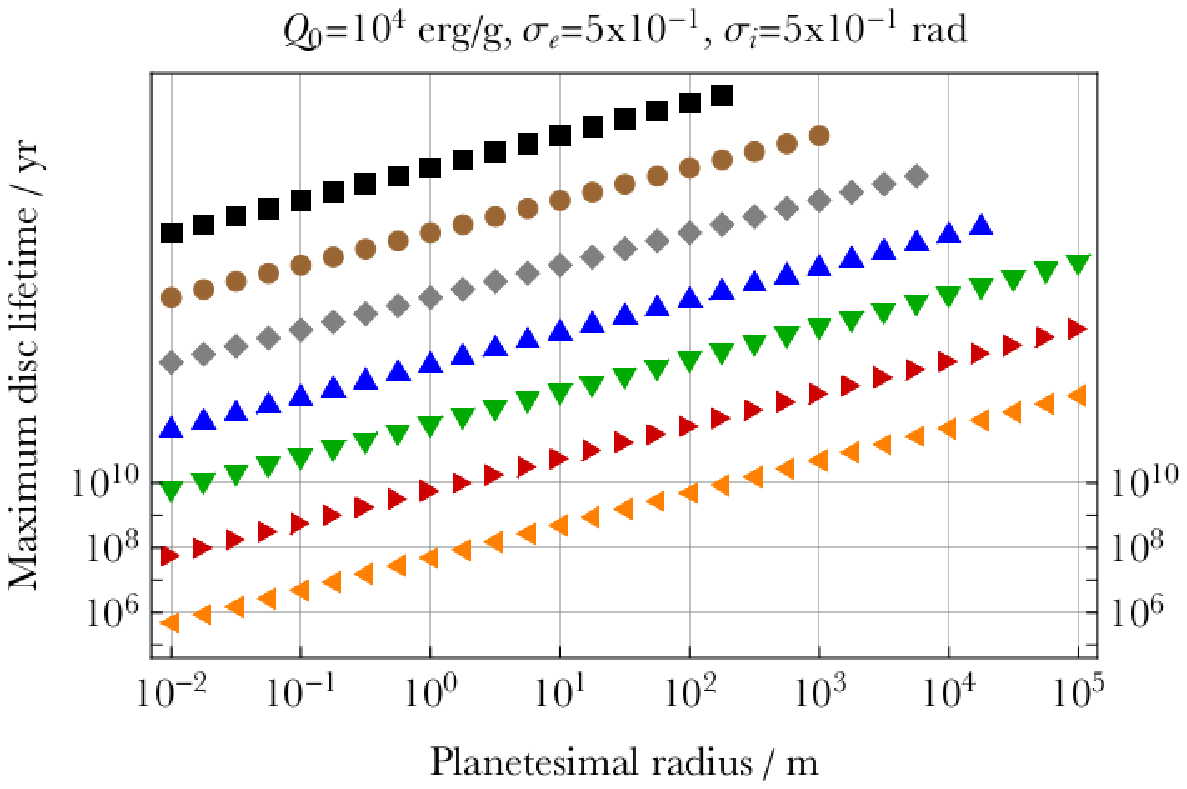}
}
\caption{
{\rev Maximum debris disc lifetimes of broad and distant ``Scattered disc''-like YORP discs (formed from YORP break up during the giant branch phase), utilizing the legend in Fig. \ref{YORP1}. The plots vary $\sigma_e$, $\sigma_i$ and $Q_0$ in different combinations, but all assume $\rho_{\rm p}=2$~g/cm$^3$, $M_{\star} = 0.6M_{\odot}$, $r_{\rm i} = 30$ au and $r_{\rm o} = 100$ au. The plots demonstrate that in all cases, broad YORP discs as massive as $10^{24}$~kg live long enough to source intact planetesimals to the white dwarf.}
}
\label{YORP2}
\end{figure*}

\begin{figure*}
\centerline{ {\bf \LARGE \underline{Spin debris discs from $1.5-4.5R_{\odot}$}} }
\centerline{}
\centerline{}
\centerline{
\includegraphics[width=9.5cm]{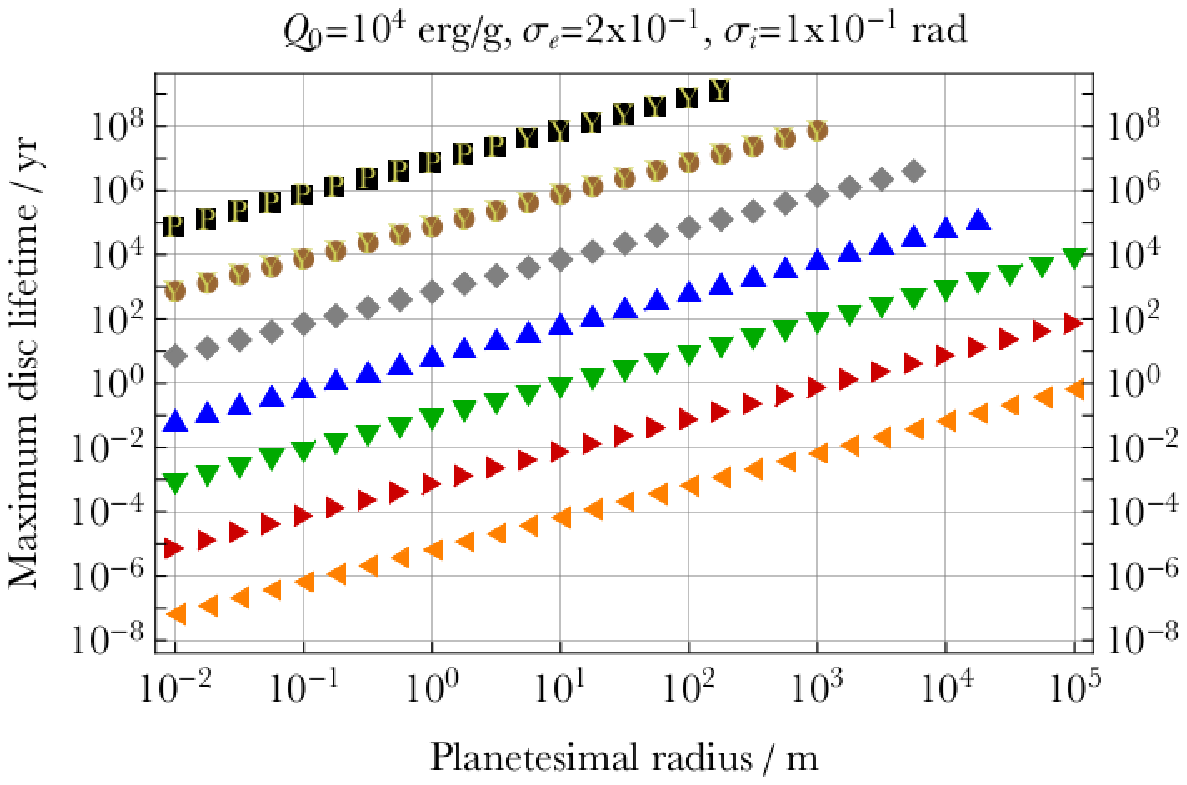}
\ \ \ \ \ \
\includegraphics[width=9.5cm]{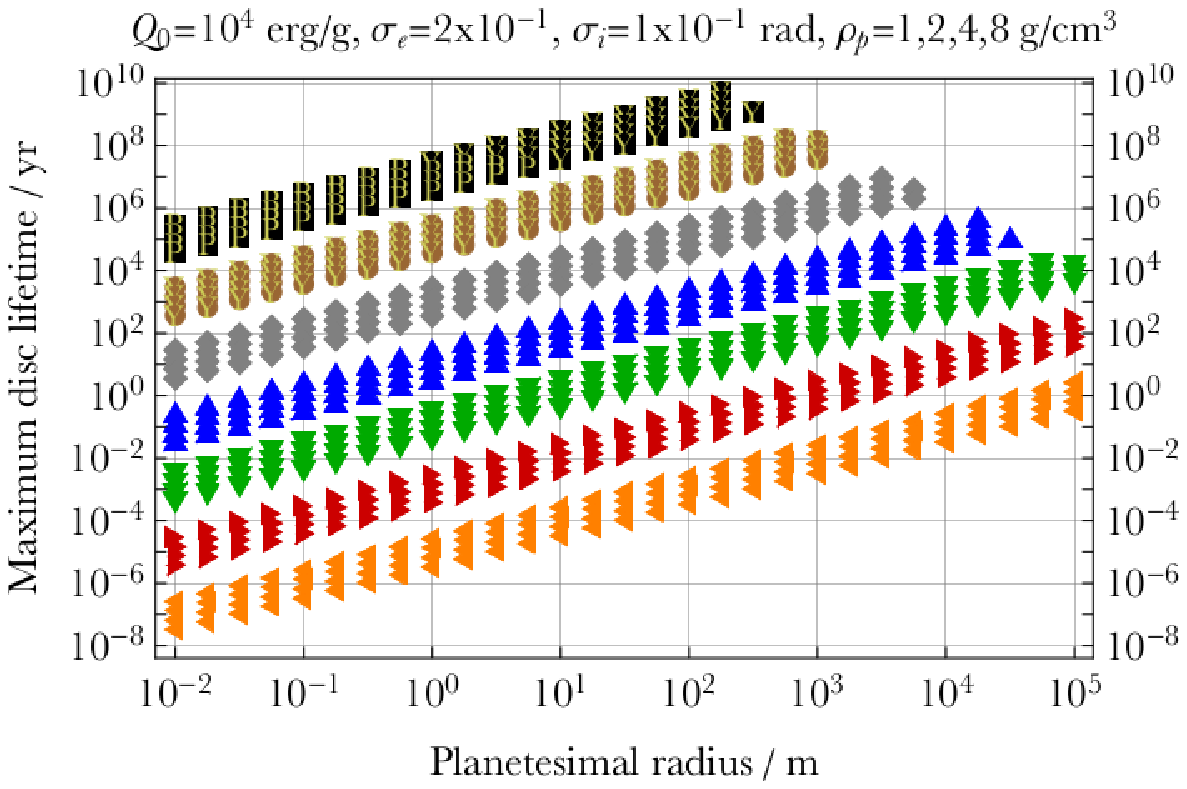}
}
\centerline{}
\centerline{
\includegraphics[width=9.5cm]{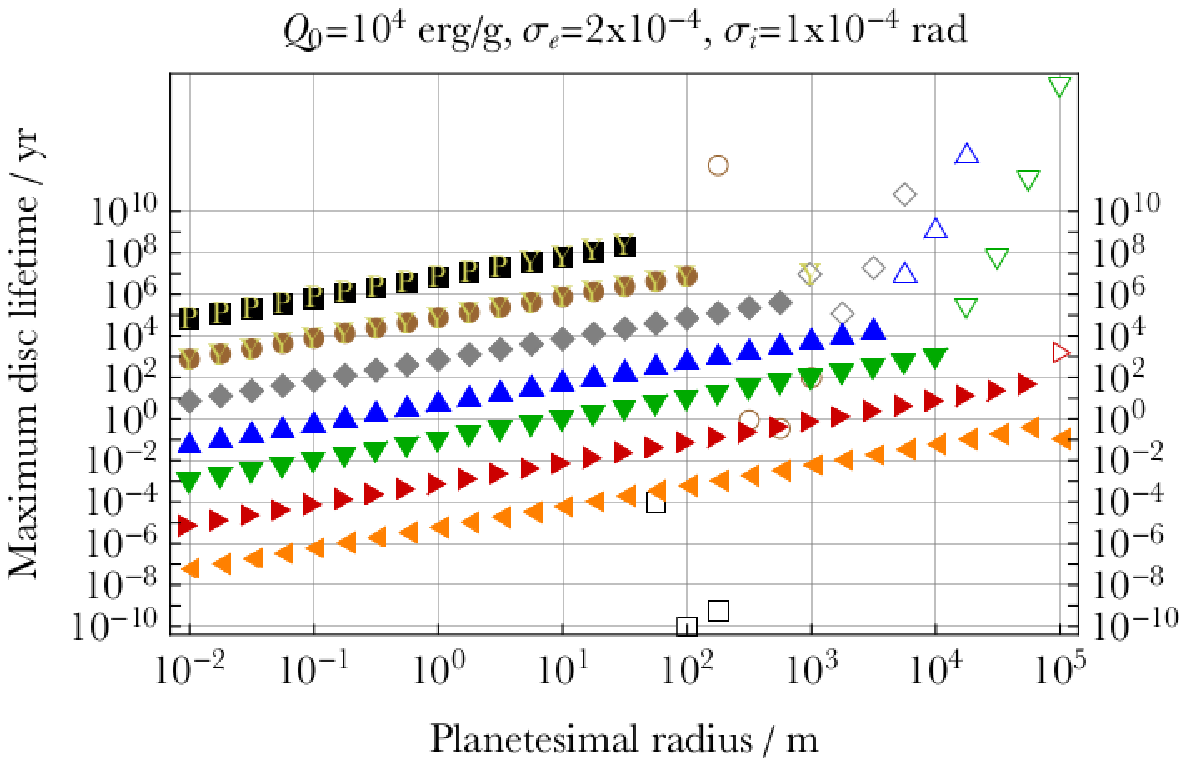}
\ \ \ \ \ \
\includegraphics[width=9.5cm]{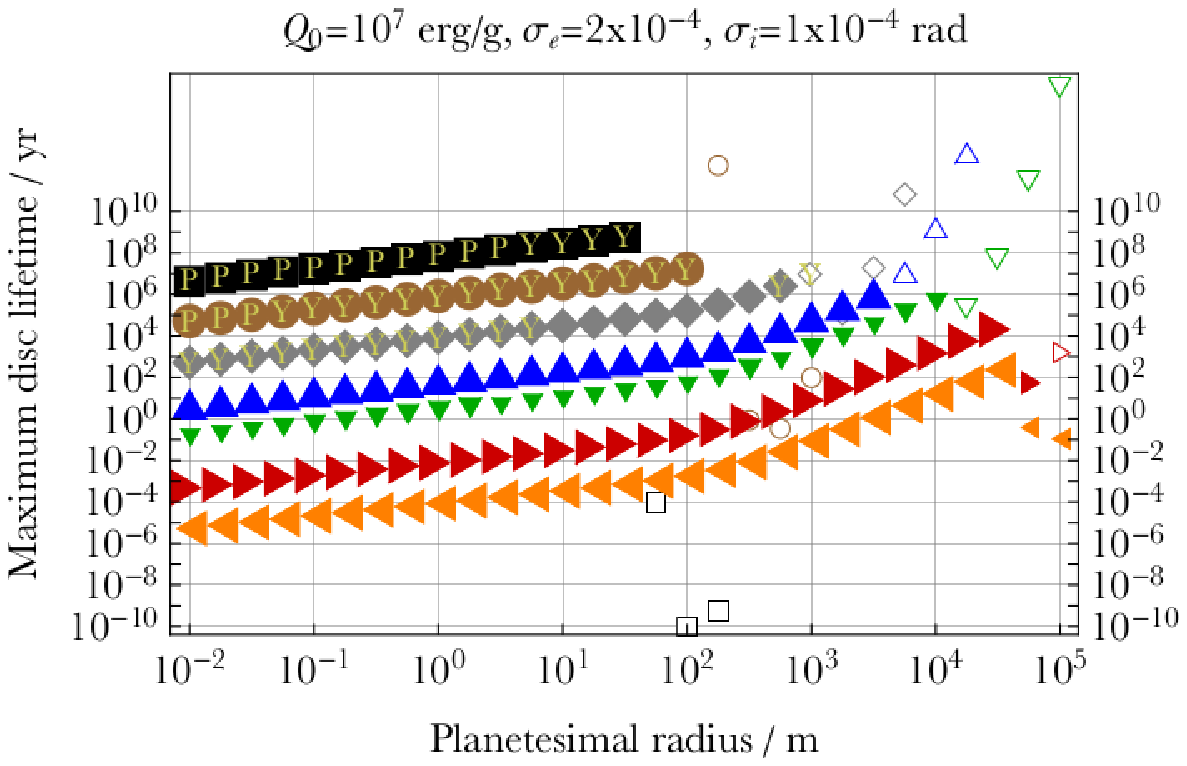}
}
\caption{
{\rev 
Maximum debris disc lifetimes of spin discs (formed from radiation-less rotational fission during the white dwarf phase), utilizing the legend in Fig. \ref{YORP1}.  For all plots, $M_{\star} = 0.6M_{\odot}$, $r_{\rm i} = 1.5 R_{\odot}$ and $r_{\rm o} = 4.5 R_{\odot}$. The upper right plot quantifies the (small) effect of varying $\rho_{\rm p}$; the bottom symbols correspond to $\rho_{\rm p}=1$~g/cm and the top symbols correspond to $\rho_{\rm p}=8$~g/cm. In all other plots, $\rho_{\rm p}=2$~g/cm. The plots illustrate that the maximum lifetimes of spin discs can vary from a near instantaneous breakup to about 1 Myr for young, luminous white dwarfs ($0.1L_{\odot}$), depending on both $M_{\rm disc}$ and $R_{\rm p}$. Dimmer white dwarfs might admit longer lifetimes when $M_{\rm disc} \lesssim 10^{16}$ kg.
}
}
\label{Spin}
\end{figure*}

\begin{figure*}
\centerline{ {\bf \LARGE \underline{Tidal debris discs from $0.6-1.2R_{\odot}$}} }
\centerline{}
\centerline{}
\centerline{
\includegraphics[width=9.5cm]{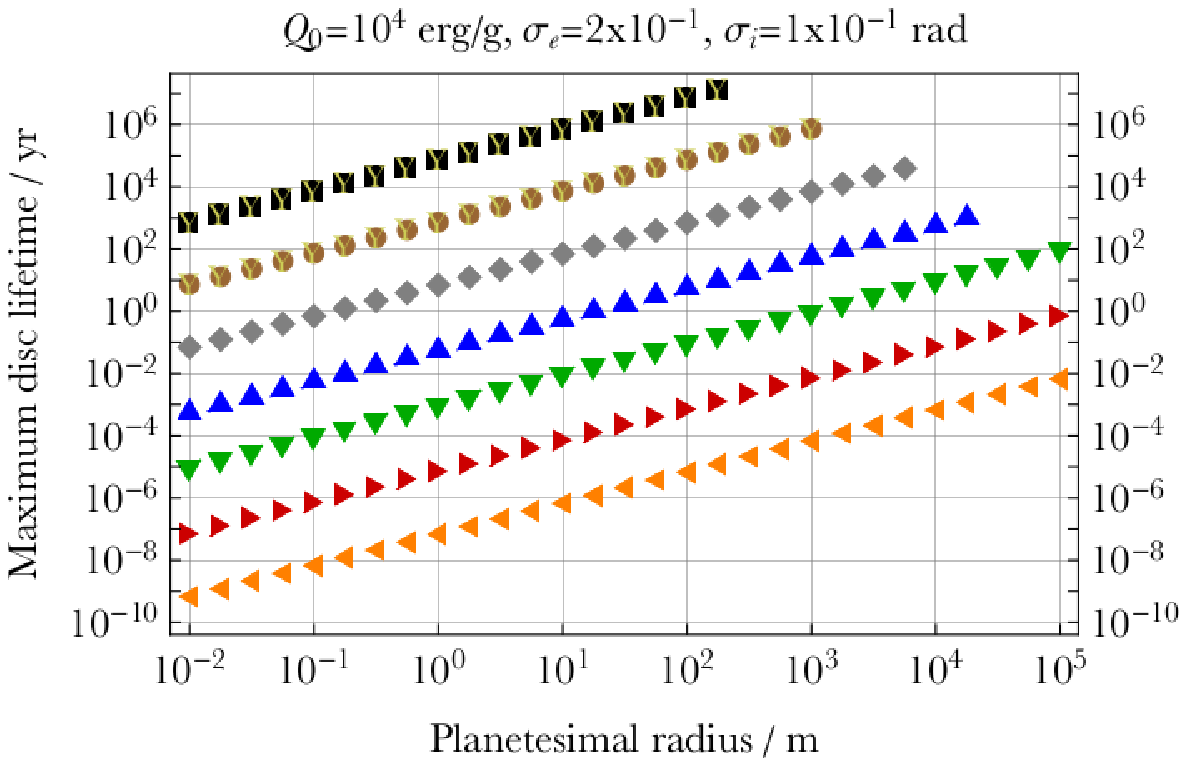}
\ \ \ \ \ \
\includegraphics[width=9.5cm]{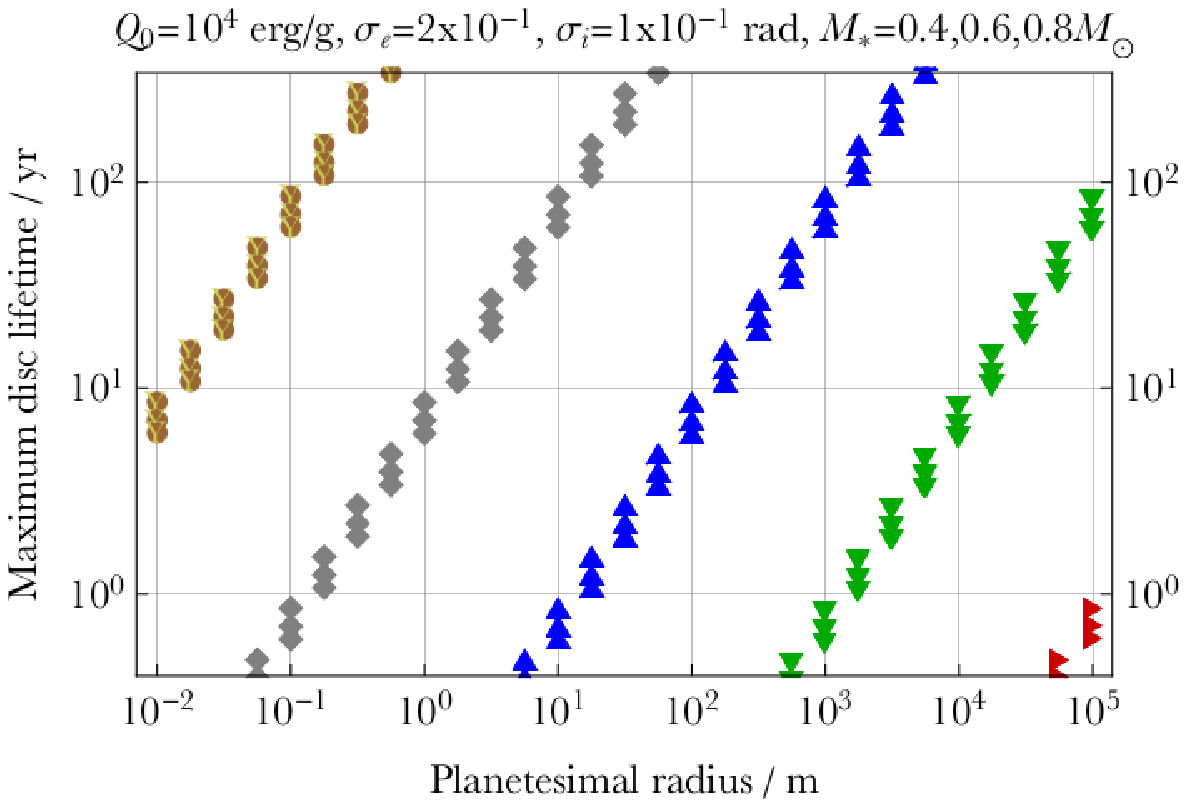}
}
\centerline{}
\centerline{
\includegraphics[width=9.5cm]{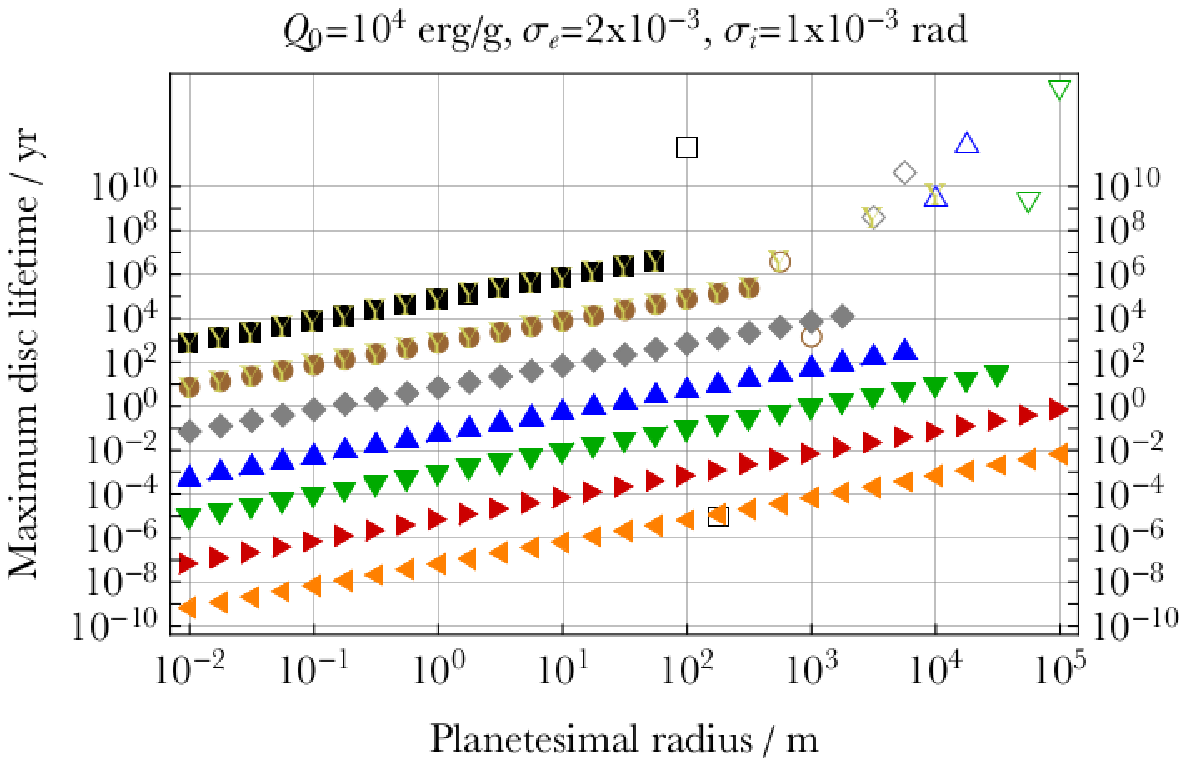}
\ \ \ \ \ \
\includegraphics[width=9.5cm]{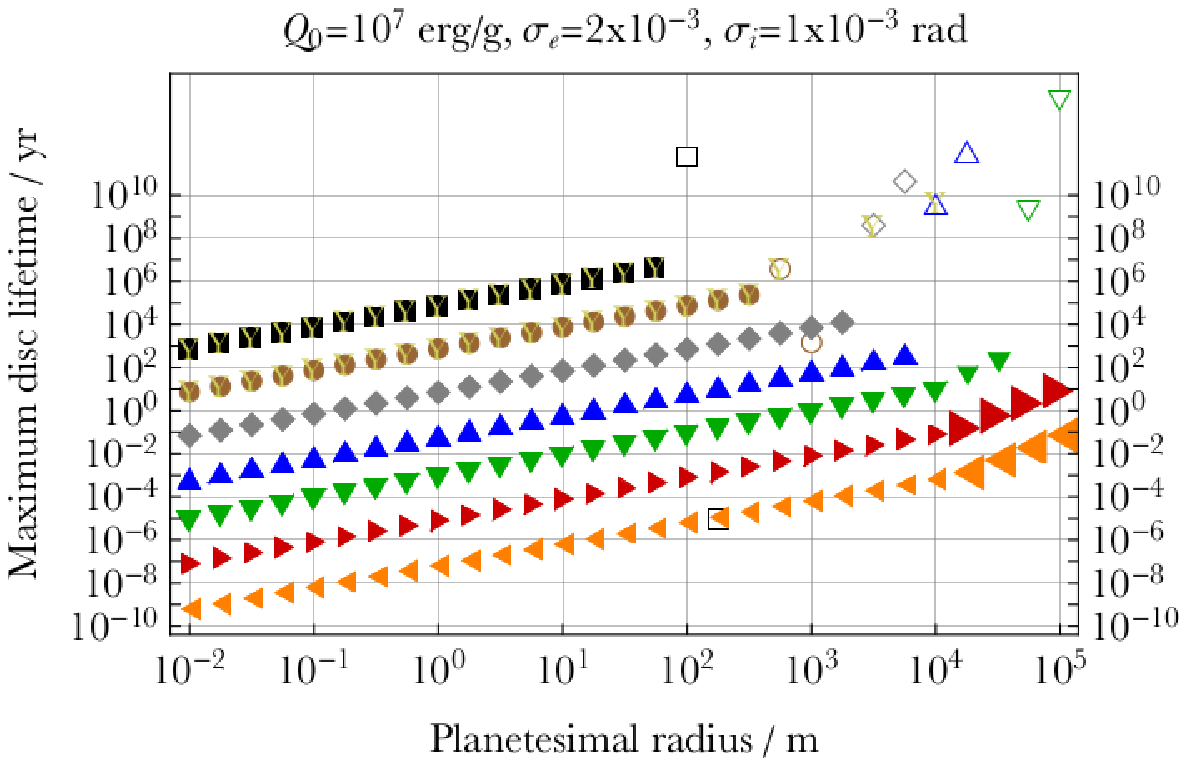}
}
\centerline{}
\centerline{
\includegraphics[width=9.5cm]{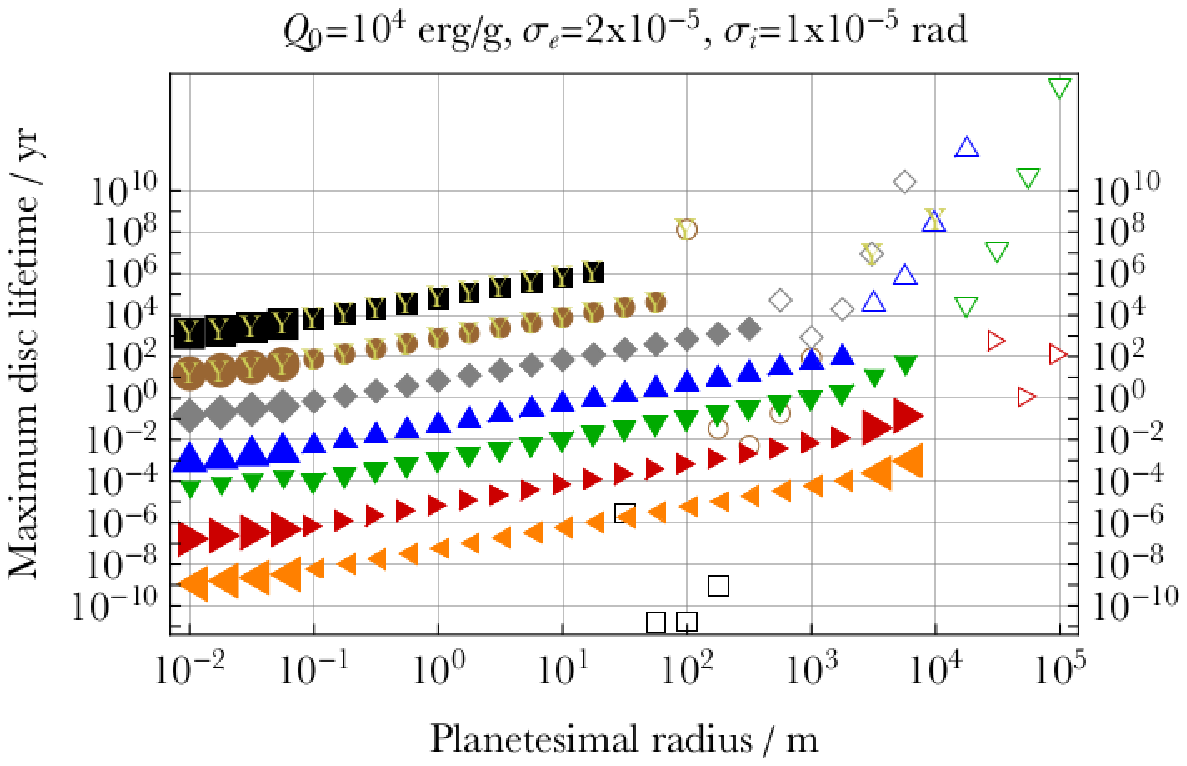}
\ \ \ \ \ \
\includegraphics[width=9.5cm]{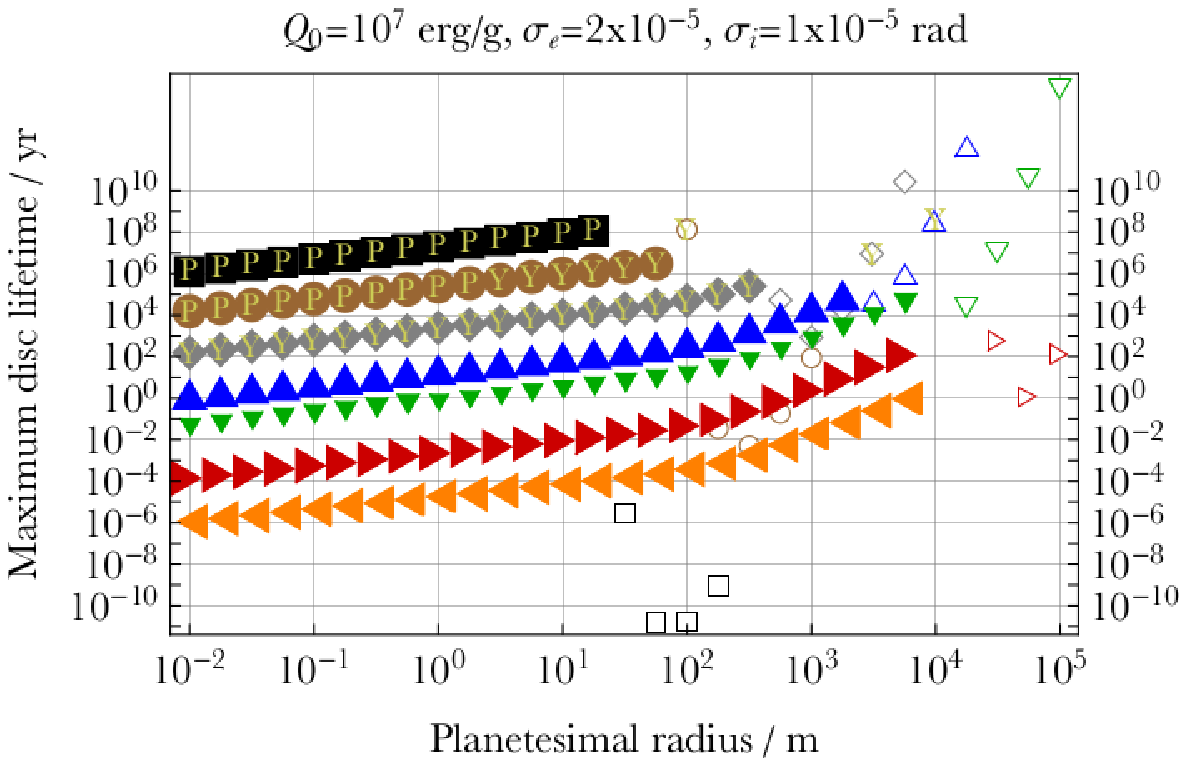}
}
\caption{
{\rev 
Maximum debris disc lifetimes of broad tidal discs (formed from destruction of bodies which travel into the white dwarf's tidal disruption distance), utilizing the legend in Fig. \ref{YORP1}. For all plots, $\rho_{\rm p}=2$~g/cm, $r_{\rm i} = 0.6 R_{\odot}$ and $r_{\rm o} = 1.2 R_{\odot}$. The upper right plot quantifies the (very small) effect of varying $M_{\star}$; the top symbols correspond to $M_{\star}=0.4M_{\odot}$ and the bottom symbols correspond to $M_{\star}=0.8M_{\odot}$. In all other plots, $M_{\star} = 0.6M_{\odot}$. The plots illustrate that the lifetimes of tidal discs are highly dependent on both $M_{\rm disc}$ and $R_{\rm p}$, and may be shorter or longer than observable timescales of years; the maximum disc lifetime is about $10^4$ yr, although some cold disc configurations may exist for much longer.
}
}
\label{Tidal1}
\end{figure*}

\begin{figure}
\centerline{ {\bf \Large \underline{Tidal debris rings}} }
\centerline{}
\centerline{}
\centerline{
\includegraphics[width=9.5cm]{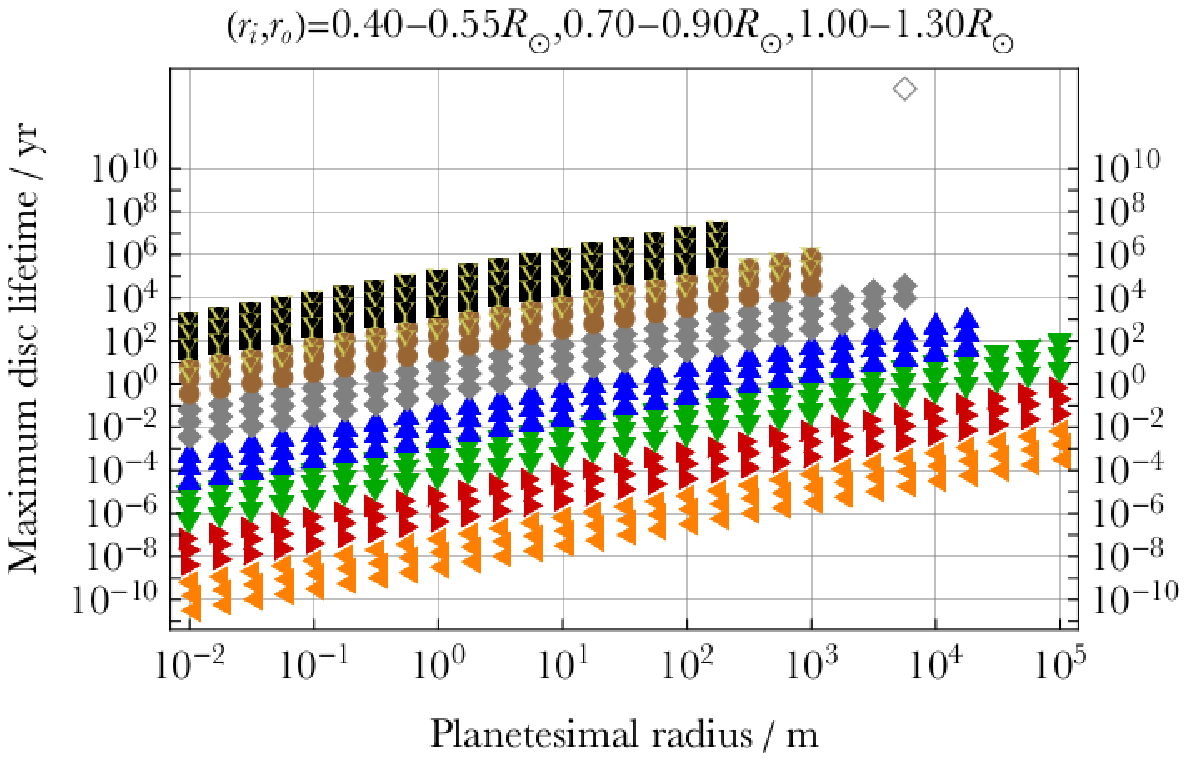}
}
\caption{
{\rev 
Maximum lifetimes of tidal debris rings (effectively narrow discs formed from destruction of bodies which enter the white dwarf's tidal disruption distance), utilizing the legend in Fig. \ref{YORP1}. Here, $M_{\star} = 0.6M_{\odot}$, $\rho_{\rm p}=2$~g/cm, and $Q_0 = 10^4$ erg/g. The upper, middle and lower sets of symbols respectively correspond to $(r_{\rm i} - r_{\rm o}) = (1.00-1.30, \ 0.70-0.90, \ 0.40-0.55)R_{\odot}$. The plot illustrates that the maximum lifetimes of tidal debris rings can vary by two orders of magnitude depending on the location of the ring within the tidal zone.
}
}
\label{Tidal2}
\end{figure}

We now {\rev compute max($t_{\rm disc}$)} for ensembles of the three types of debris discs that are illustrated in Fig. \ref{cart}: YORP-generated discs from the giant branch phases which now orbit white dwarfs (Figs. \ref{YORP1}-\ref{YORP2}), discs generated from rotational fission during the white dwarf phase (Fig. \ref{Spin}), and discs generated from tidal disruption during the white dwarf phase (Fig. \ref{Tidal1}-\ref{Tidal2}). A preliminary analysis of the parameter space revealed that the most useful and illustrative manner in which to present the results is with plots of {\rev max($t_{\rm disc}$)} versus $R_{\rm p}$, for different curves of $M_{\rm disc}$.

{\rev The legend for all of the plots in Figs. \ref{YORP1}-\ref{Tidal2} is at the top of Fig. \ref{YORP1}. In all plots, maximum lifetimes were computed for the 7 different $M_{\rm disc}$ values in the legend and for 29 uniformly spaced $R_{\rm p}$ values from 1 cm to 100 km. As indicated in the legend, the absence of a symbol indicates that the corresponding combination of $(R_{\rm p}, M_{\rm disc})$ either does not satisfy the condition for a disc to exist (Eq. \ref{disc10}) or creates a Toomre unstable disc. In order to display all symbols that provide maximum lifetime estimates, the $y$-axes of all plots were extended as necessary, although $10^{10}$ yr is the highest-valued label that was used to indicate the approximate age of the universe.}

\subsection{YORP discs}


We present results for broad (Fig. \ref{YORP1}) and narrow (Fig. \ref{YORP2}) YORP discs, {\rev roughly corresponding to analogues of the solar system's Main Belt and Scattered Disc. Both plots illustrate that the nature of the discs and the maximum disc lifetime does vary depending on how flat and circular the planetesimal orbits are (through $\sigma_e$ and $\sigma_i$). Both the hot discs and the discs labelled as ``hot or warm'' exhibit a clear correlation between max($t_{\rm disc}$)  and $R_{\rm p}$ for sub-m and often sub-km planetesimals. For larger planetesimals, the total number of planetesimals in the disc is fewer, cold discs become a possibility, and hot discs follow a different pattern.} 

{\rev For the narrow 2-3 au YORP discs in Fig. \ref{YORP1}, the profiles of the symbol sets become more complex as the disc flattens and becomes more circular. In the most excited case, where $\sigma_e = 2 \times 10^{-2} = 2 \sigma_{i}$, all discs are hot and follow linear trends in logarithmic space for max($t_{\rm disc}$) as a function of $R_{\rm p}$. All discs with $M_{\rm disc} \le 10^{14}$ kg could survive for the age of the universe, as could some with $M_{\rm disc} = 10^{20}$ kg. Even a few $M_{\rm disc} = 10^{24}$ kg discs could survive for up to 10 Myr. More flattened, calm discs may be hot, warm or cold. In no case are the disc lifetimes limited by Poynting-Robertson drag, although $10^{16-20}$ kg discs composed of sub-m sized planetesimals may be influenced by the white dwarf luminosity.
}

{\rev 
The broader 30-100 au YORP discs in Fig. \ref{YORP2} are all too far away to be affected by white dwarf radiation. Their maximum lifetimes are also much higher than in the Main Belt analogue case for sub-km planetesimals and often for larger planetesimals. The six plots in the figure vary $\sigma_e$, $\sigma_i$ and $Q_0$ to demonstrate the complexity that this variation induces, particularly for the discs composed of the fewest number of planetesimals. The upper two panels are plotted within the same ranges and with the only difference being that $Q_0$ varies by three orders of magnitude. The same is true for the middle two panels. These four panels together show that $Q_0$ can vary maximum disc lifetimes for ``hot or warm'' discs only, and by up to three orders of magnitude. 

The most excited of these disc sets, with $\sigma_e = \sigma_i = 0.5$, is shown in the bottom right plot. Here, all discs are hot and the only discs which do {\it not} survive for at least 10 Gyr are sub-km discs with $M_{\rm disc} \ge 10^{22}$ kg. If $\sigma_e \ll \sigma_i$, then the result is shown in the bottom left plot. Comparison with the middle left plot ($\sigma_e = \sigma_i$) showcases only minor differences: only one order of magnitude in max($t_{\rm disc}$), and for hot discs only.
}

\subsection{Spin discs}

The spatial scale of {\rev spin} discs is orders of magnitude smaller than {\rev that of} YORP discs, but is beyond the tidal radius of the white dwarf. The rough and under-investigated outer boundary of $4.5R_{\odot}$ {\rev reduces the disc lifetime significantly compared to YORP discs, as shown in Fig.~\ref{Spin}. Spin discs are also more prone to disruption from white dwarf radiation.
}

The spatial scales of both {\rev spin} and tidal discs are comparable to Saturn's rings, which can be near-circular down to the $\sim 10^{-4}$ eccentricity level. {\rev However, the formation channel of spin discs could be very different, and the resulting planetesimal orbits may instead feature $\sigma_e \gg 10^{-4}$ and $\sigma_i \gg 10^{-4}$. Therefore, the upper two plots of Fig. \ref{Spin} illustrate ``excited'' cases while the lower two plots instead illustrate ``calm'' cases. The only difference in the calm case plots is the value of $Q_0$: only for the higher value are warm discs possible. The upper right plot illustrates the effect of altering $\rho_{\rm p}$; across the entire range of planetesimal density, the maximum disc lifetime varies by less than two orders of magnitude.

All four plots demonstrate that spin discs may reside in a steady state for up to about 1 Myr. However, discs with masses under about $10^{16}$ kg could be subject to dominant effects from highly luminous ($0.1L_{\odot}$) white dwarf radiation, and those with $M_{\rm disc} \approx 10^{12}$~kg are often disrupted first by Poynting-Robertson drag. 
}

\subsection{Tidal discs}

{\rev Spin} debris discs may reside in locations {\rev which are} up to an order-of-magnitude more distant than tidal debris discs. We explore how such a spatial difference affects max($t_{\rm disc}$) for tidal debris discs in Figs. \ref{Tidal1}-\ref{Tidal2}, but first provide a reminder about the assumptions in our model. We are assuming that the disc is gas-free, and hence further away than the sublimation distance, which directly relates to the cooling age (see Fig. \ref{cart}). Further, for warm and hot discs, any gas produced through collisions is assumed to be negligible. Finally, we assume that the planetesimals themselves, once settled into a disc, do not break apart due to tidal forces.

{\rev The overall message conveyed by both Figs. \ref{Tidal1} and \ref{Tidal2} is that the maximum lifetimes of tidal discs is $\sim 10^4$~yr, but may be much shorter, on observable timescales of years. Figure \ref{Tidal1} presents broad tidal discs ranging from $0.6-1.2R_{\odot}$. In the figure, the left panels moving downward feature increasingly flattened and circular tidal discs. The consequence is an increase in the resulting variation of disc types (hot, warm and cold) but a marginal change in the maximum disc lifetimes of the hot and warm discs. As shown in the upper right panel, the dependence of maximum lifetime on $M_{\star}$ is negligible. The other plots in the right panels show discs for which the same initial conditions were adopted as in the left panels, except for a change in the value of $Q_0$. Increasing $Q_0$ allows for more warm discs to occur, and increases the maximum disc lifetimes by several orders of magnitude.}

{\rev Figure \ref{Tidal2} instead explores the case where tidal disruption produces ring-like instead of disc-like structures \citep{debetal2012,veretal2014a,malper2020a,malper2020b}. The plot illustrates the variation in the maximum debris disc lifetime due to different placements of narrow discs within $1.3R_{\odot}$ of the star. This location can vary the disc lifetimes by two orders of magnitude (the further away the ring, the higher the maximum lifetime). The innermost regions are only plausible in the first place for the oldest white dwarfs, where the disc avoids sublimation (see Fig. \ref{cart}).}

\section{Discussion}

\subsection{Implications of results}

The implications of our results vary depending on which of the three types of discs is the focus of discussion, {\rev although the origin and evolution of all three discs are connected}. YORP discs represent an important, if not the primary, source of mass which eventually is perturbed towards and pollutes the white dwarf. Whether or not this mass is already ground down into dust before the perturbation can help determine the rarity of objects like those orbiting WD 1145+017 and ZTF J0139+5245. The mass of the progenitor of the debris orbiting WD 1145+017 is thought to contain about $10^{20}$ kg of mass \citep{rapetal2016,veretal2017a,guretal2017} whereas the mass of the progenitor in the ZTF J0139+5245 system is not as well constrained.

Figures \ref{YORP1}-\ref{YORP2} illustrate {\rev definitively} that YORP discs {\rev are a viable reservoir which may supply planetesimals to the white dwarf for cooling ages exceeding $10$ Gyr. This result is independent of planetesimal size, which the plots demonstrate could span seven orders of magnitude from 1 cm to 100 km. Although both plots illustrate unlikely cases where the discs may be cold, the majority are most likely hot. Hot discs are common when $\sigma_e, \sigma_i > 10^{-3}$, and we expect YORP discs to form in dynamically excited states. 

The reason is because the giant branch Yarkovsky effect acts concurrently on progenitor asteroids and will have probably already significantly altered their orbital eccentricity and inclination \citep{veretal2015a,veretal2019a} before YORP breakup. The typical fragmentation sizes for YORP breakup are unknown, and this process around giant branch stars has yet to be modelled in detail. \cite{versch2020} did, however, establish that multiple fission events may easily occur in quick succession during the tip of the asymptotic giant branch phases for Main Belt analogues. If this process is common, then we might expect YORP discs to be composed of small (cm-sized or m-sized) monoliths of high internal strength. In Scattered Disc analogues, for which multiple fissions would be less common, the fragmented bodies are more likely to be km-sized rubble piles.

If two terrestrial planets collide with one another at au-scales during the white dwarf phase, the resulting fragments (or planetesimals) could form a different type of disc for which our YORP disc computations would still be applicable. Another possibility is that extant planetesimals could collide with existing terrestrial planets. The resulting impact ejecta may be thrust towards, perturbed close to, or radiatively dragged towards the white dwarf \citep{verkur2020}. Regardless of how the planetesimals are formed, they would be subject to gravitational perturbations from surviving major planets \citep{debsig2002,bonetal2011,debetal2012,veretal2013,voyetal2013,frehan2014,vergan2015,antver2016,antver2019,veretal2016a,musetal2018,smaetal2018}.

Observing the transfer of matter from a YORP disc to a spin or tidal disc is challenging. Current capabilities are limited to detecting dust \citep{xuetal2013,faretal2014}, although the highly eccentric transiting material (with an eccentricity of 0.97) orbiting ZTF J0139+5245 \citep{vanetal2019} may be indicative of a transfer-in-progress.  This progenitor of the observed debris could have broken up due to either rotational fission or tidal disruption depending on the location of its orbital pericentre and its physical shape. If rotational fission is the origin, then subsequent to radiative circularization \citep{veretal2015b}, the debris will form a spin disc. 

We have shown here that the survival timescales for spin discs are strongly dependent on the values of $R_{\rm p}$ and $M_{\rm disc}$, and can vary from seconds to about 1 Myr (despite some cold disc cases with small numbers of planetesimals where the survival timescales exceed this value). If the disc mass in ZTF J0139+5245 is $10^{14}-10^{15}$~kg \citep{veretal2020a}, then Fig. \ref{Spin} demonstrates that the lifetime of this disc is at least hundreds if not thousands of years; only if the star was brighter would the disc be influenced by the Yarkovsky effect.

Maximum lifetimes of tidal debris discs are a couple of orders of magnitude shorter than those of spin debris discs. Figs. \ref{Tidal1}-\ref{Tidal2} suggest that 0.01 Myr may be considered as an upper bound for tidal discs before they are sure to break down into dust and/or gas. However, our phase space exploration indicates that most of the tidal discs that we constructed would depart from a steady state within observable timescales of years\footnote{{\rev The standout exceptions are the cold tidal discs with small numbers of large planetesimals. These discs are physically plausible given that, at least in WD 1145+017, the orbital eccentricity of the progenitor asteroid is probably under $1 \times 10^{-2}$ \citep{guretal2017,veretal2017a}}}.
}

Consequently, detecting photometric curves of just planetesimals -- without any broad tails due to dust -- would provide particularly valuable evolutionary constraints and spur additional imminent observations.  In the case of WD 1145+017, the disc contains gas, dust and fragments, such that the likely lone progenitor object in that system cannot be thought of to reside in the types of debris discs that we consider here. 

However, \cite{veretal2016b} investigated gas-free disc analogues of the WD 1145+017 planetary system with 4, 6 and 8 planetesimals with $M_{\rm p} = 10^{17}-10^{23}$ kg. They ran $N$-body simulations lasting five years primarily to determine transit timing variations, but also secondarily to determine the instability timescales of these debris discs. Over their simulation duration, their $M_{\rm disc} = 4-8\times 10^{17}$ kg discs remained stable.  However, their study is not necessarily comparable {\rev to ours} because they placed the planetesimals on exactly circular and co-orbital orbits ($\sigma_e = \sigma_i = 0$). As we have demonstrated, small eccentricity dispersions on the order of $10^{-5}$ can make a significant difference to {\rev max($t_{\rm disc}$)}. 

{\rev Our computations for max($t_{\rm disc}$) assume that the disc is not influenced by external forces. External accretion onto the disc itself from incoming material \citep{broetal2017} may be common and affect the steady state lifetime \citep{kenbro2017a}. Further, the timescale on which major planets perturb planetesimals from the YORP disc close to the white dwarf may be shorter than max($t_{\rm disc}$). However, this perturbation timescale is highly model dependent.

Another type of external influence on the disc itself are these major planets. Tidal migration of these planets would disrupt if not destroy extant spin or tidal discs. High-eccentricity \citep{ocolai2020} and chaotic \citep{verful2019} migration could see a planet plough through a white dwarf disc.  Slower tidal migration mechanisms that rely on magnetism \citep{broken2019,verwol2019} and quality function variations \citep{veretal2019b} could also represent disruptive influences, just on a different timescale.  

Even neglecting potential external influences on the discs, one may attempt to link $M_{\rm disc}$ and max($t_{\rm disc}$) for YORP, spin and tidal discs with observed accretion rates onto white dwarf atmospheres. However, there are three crucial missing pieces to this story: the delivery frequency of YORP planetesimals to the vicinity of the white dwarf, the amount of accreted material which bypasses the disc phase entirely \citep{wyaetal2014,broetal2017,turwya2020}, and the timescale for the sublimated gas to accrete onto the white dwarf. These quantities are highly model dependent, and the last is likely to require the application of sophisticated simulations with a coupled treatment of gas and dust \citep{metetal2012,kenbro2017b}.
}

\subsection{Collision-limited discs}

\begin{figure}
\centerline{ {\bf \Large \underline{Collision-limited hot tidal discs}} }
\centerline{
\includegraphics[width=9.0cm]{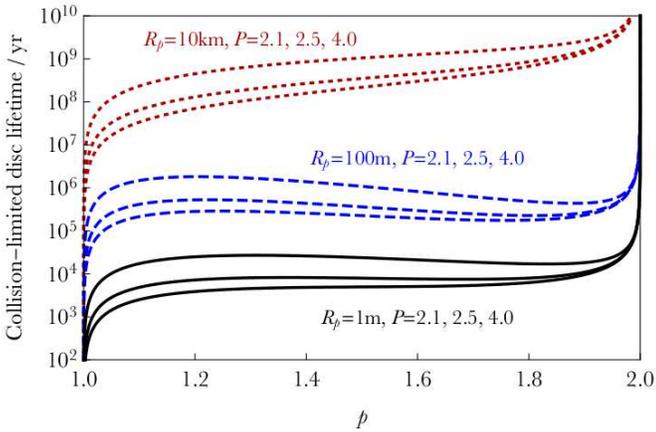}
}
\caption{
{\rev 
Lifetimes of collision-dominated hot tidal discs [$(r_{\rm i} - r_{\rm o})=(0.6 - 1.2)R_{\odot}, \sigma_e = 2 \times 10^{-3} = 2\sigma_i, Q_0 = 10^{7}$ erg/g, $M_{\rm disc} = 10^{20}$ kg, $\rho_{\rm p} = 2$ g/cm$^3$, $M_{\star} = 0.6M_{\odot}$] which are not monodisperse, but rather follow a power-law mass distribution given by Eq. (\ref{sizedist}). Plotted are different exponents ($p,P$) for this distribution, as well as three different values of the planetesimal radius $R_{\rm p}$ corresponding to the maximum planetesimal mass. 
}
}
\label{Collision1}
\end{figure}

{\rev

Our investigation has been limited in many respects so that we could focus on computing steady state debris disc lifetimes. Two of our most significant limitations are (i) our assumption that the discs are monodisperse, and that (ii) our computations do not include the disc's subsequent evolution after leaving the steady state. This evolution would require a dedicated detailed investigation \citep{bocraf2011,rafikov2011a,rafikov2011b,metetal2012,rafgar2012,kenbro2017a,kenbro2017b,mirraf2018} and preferably include the physical processes of fragmentation, erosion, collisions, sublimation, condensation and ultimately accretion onto the white dwarf. 

Relaxing the assumption of equal-mass planetesimals would require a different model, one which might not admit a tractable analytical form as in \cite{hentre2010}. However, with the existing model, we can still obtain useful results for dispersion-dominated, hot discs by assuming that they are ``collision-limited'' with a power-law -- rather than a monodisperse -- mass distribution. The reason is because an initial power-law mass distribution which produces a collision-limited disc that is dominated by small bodies settles towards a state which mimics a monodisperse system of fragments with mass $M_{\rm max}$.

Section 4.3 of \cite{hentre2010} presents a method to compute $M_{\rm max}$ as an implicit function of time. We have a different goal here of computing $t_{\rm disc}$. We slightly generalize their treatment by starting with Eq. (\ref{dispreg}), from which we can now write

\[
\left(t_{\rm disc}\right)^{-1} = 
27.68 \frac{\sigma_r \sigma_i}{\sigma_e} 
\left( \frac{3M_{\rm max}}{4 \pi \rho_{\rm p}} \right)^{\frac{2}{3}}
\int_{M_{\rm min}}^{M_{\rm max}} \frac{dn_0(M_{\rm p}')}{dM_{\rm p}'} dM_{\rm p}' 
.
\]

\begin{equation}
\end{equation}

Further, assume that the initial mass distribution in the disc is given by the following broken power-law

\[
\frac{dn_0(M_{\rm p})}{dM_{\rm p}} = \frac{n_{\rm max}}{M_{\rm max}} \left( \frac{M_{\rm p}}{M_{\rm max}} \right)^{-p}, \ \ \ \  p < 2 \ {\rm and}
 \ M_{\rm p} \le M_{\rm max}
\]

\[
\frac{dn_0(M_{\rm p})}{dM_{\rm p}} = \frac{n_{\rm max}}{M_{\rm max}} \left( \frac{M_{\rm p}}{M_{\rm max}} \right)^{-P}, \ \ \   P > 2 \ {\rm and} 
 \ M_{\rm p} \ge M_{\rm max}
\]

\begin{equation}
\label{sizedist}
\end{equation}

\noindent{}where

\begin{equation}
n_{\rm max} = \frac{\rho_0}{M_{\rm max}} \left( \frac{1}{2-p} + \frac{1}{P-2}  \right)^{-1} 
\end{equation}

\noindent{}and where the minimum and maximum planetesimal masses in the disc are related through

\begin{equation}
M_{\rm min} = M_{\rm max} \frac{Q_{\rm D}^{\star}}{\sigma_{r}^2}
.
\end{equation}

\noindent{}Integration finally yields

\begin{equation}
\left(t_{\rm disc}\right)^{-1} = 
27.68 \frac{\sigma_r \sigma_i}{\sigma_e} 
\left( \frac{3M_{\rm max}}{4 \pi \rho_{\rm p}} \right)^{\frac{2}{3}}
\frac{n_{\rm max}}{p-1}
\left( \frac{Q_{\rm D}^{\star}}{\sigma_{r}^2} \right)^{1-p}
\end{equation}

\noindent{}where $Q_{\rm D}^{\star}$ is a function of $M_{\rm max}$ through Eq. (\ref{Qeq}). We can now compute disc lifetime as a function of the steepness of the size distribution, and do plot the result in Fig. \ref{Collision1}. The plot illustrates that a uniform decrease in $R_{\rm p}$ does not correspond to a uniform decrease in $t_{\rm life}$, and that $t_{\rm life}$ is relatively independent of $P$ when $P > 2.5$.

}

\section{Summary}

The evolution timescales of planetary debris discs around white dwarfs represent crucial parameters for our understanding of post-main-sequence planetary science. Here, we analytically computed the {\rev maximum steady state} lifetimes of three types of {\rev monodisperse} white dwarf debris discs: those composed of debris from (i) giant branch YORP break-up at $2-100$ au, (ii) radiation-less rotational fission at $1.5-4.5R_{\odot}$, and (iii) tidal disruption within $1.3R_{\odot}$. We displayed a series of figures which cover nearly the entire relevant parameter space, and hence can be used to read off and extrapolate {\rev bounds} for individual systems or ensembles. 

{\rev We found that YORP discs of masses ranging from $10^{12}-10^{24}$~kg are sufficiently long-lived to provide a reservoir of planetesimals of sizes $10^{-2} - 10^5$~m which can be delivered {\it intact} to white dwarfs of any cooling age. The other two types of debris discs which are formed closer to the vicinity of the white dwarf cannot survive in a steady state for longer than about 1 Myr (for the fissional spin discs) and 0.01 Myr (for the tidal discs). However, in the majority of parameter space, these discs leave their steady state within about 1 yr. Hence, transit detections of white dwarf planetesimals without dusty tails should be monitored regularly to detect imminent dynamical activity.
}

\section*{Acknowledgements}

{\rev We thank the reviewer for their helpful comments, which have improved the manuscript.}
DV gratefully acknowledges the support of the STFC via an Ernest Rutherford Fellowship (grant ST/P003850/1). KH acknowledges partial financial support from the Center for Space and Habitability (CSH), the PlanetS National Center of Competence in Research (NCCR), the Swiss National Science Foundation, the MERAC Foundation and a European Research Council (ERC) Consolidator Grant (number 771620).

\label{lastpage}
\end{document}